\documentclass[nohyper]{JHEP3}

\usepackage{graphicx}
\usepackage{epsfig,cite}

\newcommand{\be}{\begin{equation}}
\newcommand{\ee}{\end{equation}}
\newcommand{\bea}{\begin{eqnarray}}
\newcommand{\eea}{\end{eqnarray}}
\newcommand{\bi}{\begin{itemize}}
\newcommand{\ei}{\end{itemize}}
\newcommand{\ben}{\begin{enumerate}}
\newcommand{\een}{\end{enumerate}}
\newcommand{\la}{\left\langle}
\newcommand{\ra}{\right\rangle}
\newcommand{\lc}{\left[}
\newcommand{\rc}{\right]}
\newcommand{\lp}{\left(}
\newcommand{\rp}{\right)}

\newcommand{\aq}{\alpha_s\left( Q^2 \right)}
\newcommand{\amz}{\alpha_s\left( M_Z^2 \right)}
\newcommand{\aqq}{\alpha_s \left( Q^2_0 \right)}

\def\toone#1{\mathrel{\mathop{\sim}\limits_{\scriptscriptstyle
{#1\rightarrow1 }}}}
\def\frac#1#2{{{#1}\over {#2}}}
\def\gsim{\mathrel{\rlap{\lower4pt\hbox{\hskip1pt$\sim$}}
    \raise1pt\hbox{$>$}}}         
\def\lsim{\mathrel{\rlap{\lower4pt\hbox{\hskip1pt$\sim$}}
    \raise1pt\hbox{$<$}}}         
\newcommand{\mrexp}{\mathrm{exp}}
\newcommand{\dat}{\mathrm{dat}}
\newcommand{\one}{\mathrm{(1)}}
\newcommand{\two}{\mathrm{(2)}}
\newcommand{\art}{\mathrm{art}} 
\newcommand{\rep}{\mathrm{rep}}
\newcommand{\net}{\mathrm{net}}

\newcommand{\sys}{\mathrm{sys}}

\newcommand{\pdf}{\mathrm{pdf}}
\newcommand{\tot}{\mathrm{tot}}
\newcommand{\minn}{\mathrm{min}}
\newcommand{\mut}{\mathrm{mut}}

\newcommand{\NS}{\mathrm{NS}}

\newcommand{\gen}{\mathrm{gen}}

\newcommand{\parr}{\mathrm{par}}

\newcommand{\extra}{\mathrm{extra}}
\newcommand{\draft}[1]{}

\title{Neural network determination of  parton distributions:
the nonsinglet case}

\author{{\bf  The NNPDF Collaboration:}\\
Luigi Del Debbio$^1$, Stefano
  Forte$^2$, Jos\'e I. Latorre$^3$, Andrea Piccione$^{4}$ and 
Joan Rojo$^{3,5}$

~$^1$ SUPA, School of Physics, University  of Edinburgh, Edinburgh EH9 3JZ, Scotland\\
~$^2$ Dipartimento di Fisica, Universit\`a di Milano and\\ 
INFN, Sezione di Milano, Via Celoria 16, I-20133 Milano, Italy\\
~$^3$ Departament d'Estructura i Constituents de la Mat\`eria, \\
Universitat de Barcelona, Diagonal 647, E-08028 Barcelona, Spain\\
~$^4$ INFN, Sezione di Genova, via Dodecaneso 33, I-16146 Genova,  Italy\\
~$^5$ LPTHE, CNRS UMR 7589\\ Universit\'es Paris VI-Paris VII,
F-75252, Paris Cedex 05, France\\

}

\abstract{We provide a determination of the isotriplet quark
  distribution from available deep--inelastic data using neural networks.
We give  a general introduction to the neural network
  approach to parton distributions, which provides a  solution
  to the problem of constructing a faithful and unbiased  
probability distribution of parton densities based on available
experimental information. 
 We
  discuss in detail the techniques which are necessary in order to
  construct a Monte Carlo representation of the data, to construct and
  evolve neural
  parton distributions, and to train them in such a way that the
  correct
statistical features of the data are reproduced.
We present the results of the application of this method to the
determination of the
nonsinglet quark distribution up to next--to--next--to--leading order,
and compare them with those obtained using other approaches.}


\preprint{UB-ECM-PF 06/17\\IFUM-882-FT \\ GEF-TH-01-2007}

\begin{document}

\section{Introduction}
\label{introduction}
The needs of precision physics at hadron colliders have determined a
revolution in the approach to the determination of parton
distributions functions
(PDFs) over the last few years. While the Tevatron has been
providing data for a variety of hard hadronic processes which
establish the validity and accuracy of perturbative factorization and
parton universality to a level which is now comparable to that of
precision tests of the standard model in the electroweak sector, 
LHC, now behind the corner, will require, essentially for the first
time, a precision approach to the structure of the nucleon in the
context of searches for new physics~\cite{lh1}.
Over a  decade of experimental effort, especially in deep-inelastic
experiments, 
first and foremost at the HERA collider~\cite{heralhc},
but also from neutrino beams and with muon beams on a fixed target, 
has provided us with
an unprecedented amount of information which makes such a precision
approach possible.  However, we have now reached a stage in which 
the development of new
theoretical and phenomenological analysis tools is needed~\cite{pdfrev}.

The main recent development in the determination of parton distributions 
has been the availability of sets of parton distributions with
errors~\cite{Alekhin,CTEQ,MRST}.
Previously, errors on PDFs where crudely estimated by
varying some sets of parameters, or comparing different determinations,
and generally considered to be negligible in comparison to other
sources of theoretical or experimental error. Now, the simultaneous
progress in higher-order theoretical calculations  and experimental
results has made this simple--minded approach obsolete: a {\it bona
  fide }
estimate of the error on PDFs has become necessary. 

This is a difficult problem, not only because of the usual difficulties in
accounting properly for theoretical uncertainties (such as 
higher order
perturbative corrections) which are non-gaussian and hard to estimate, and in
collecting and propagating uncertainties contained in large experimental
covariance matrices, but also because a PDF set is a set of functions, and
therefore one is faced with the problem of constructing an error --- a
probability measure --- on a space of functions~\cite{kosower}. 
This is
clearly an ill-posed problem, because one is trying to infer an infinite
amount of information from a finite set of data points: it can be made
tractable only by introducing some theoretical assumptions.

The standard approach to this problem is based on the choice of a
specific functional form: the infinite--dimensional problem of
determining a function is projected onto the finite--dimensional space
of parameters which determine the given parametrization. Errors on
PDFs are then essentially error ellipsoids in the finite--dimensional
parameter space, which at least in principle can be determined by
standard covariance matrix techniques.
Based on this approach, three sets of PDFs with errors have been
produced over the last few years~\cite{Alekhin,CTEQ,MRST}.
 The most obvious shortcoming of
this sort of approach is that the 
choice of a parametrization is potentially a source of bias. More
subtle problems are related to non-gaussian errors (either theoretical
and experimental) and incompatible data, which render a simple maximum
likelihood fit impossible or unreliable. 

These difficulties are apparent in the available sets of PDFs with
errors, especially when they are compared with each other. Indeed, the
direct determination of error bands as one--sigma contours seems to
be possible only if a limited set of data is fitted. In particular,
straightforward determination of errors as one--sigma contours 
has succeeded in a global PDF fit to deep-inelastic data
only~\cite{Alekhin},
 but as
soon as new data (specifically from the Drell-Yan process) are added
to this fit, their errors must be 
suitably rescaled~\cite{heralhc}. Global fits to all available 
data are instead
typically constructed by determining a priori a suitable ``tolerance
criterion'', by inspection of the quality of the fit to all available
data~\cite{CTEQ}. For example, one may determine the tolerance criterion by
considering the
spread of 90\% confidence intervals for various experiments, as one
moves away from the minimum of the $\chi^2$ along eigenvectors of the Hessian
matrix, and taking the envelope of the resulting ranges. In practice,
this suggests that 
a reasonable estimate of the one-sigma contours for
PDFs can be obtained by selecting an appropriately large value of 
$\Delta \chi^2$, such as $\Delta \chi^2=100$~\cite{CTEQ,MRST}. 
Even so, results
obtained using different parton sets for
relevant  physical observables, such as W+Higgs production at the LHC, 
are sometimes found to disagree within their stated error
bands~\cite{heralhc,djouadi}.  
The origin of these difficulties is most likely a combination of
several factors: inconsistencies between different pieces of
experimental data, underestimation of some experimental errors,
bias due to specific choices of parton parametrization, and non-gaussian
errors, especially in the space of parameters of individual parton
parametrizations.

These difficulties have stimulated various proposals for new
approaches to the determination of parton distributions, in particular
with the aim of
minimizing the bias related to the choice of a specific
functional form, and trying instead to reconstruct the probability
density in the full functional space of parton
distributions. 
An interesting suggestion~\cite{kosower} is to use Bayesian inference
based on the data  in order to update the existing prior knowledge of
PDFs, as summarized e.g.
by a Monte Carlo sample based on an existing parton
set. Although promising preliminary results have been obtained, no
parton set based on this approach has been published yet.

An alternative suggestion has been presented in Ref.~\cite{f2ns}. 
The basic idea  is to combine a
Monte Carlo sampling of the probability measure on the space of
functions that one is trying to determine (as in the approach of
Ref.~\cite{kosower}) with the use of neural
networks as universal unbiased interpolating functions. In a Monte
Carlo approach, 
the function with error --- the up quark distribution,
say ---  is given as a Monte Carlo sample of replicas of the function, 
so that any
statistical property of the underlying distribution can be derived
from the given sample. For example, the average value of the function
at some point is simply found as the average over the replicas, 
its error as the
variance and so on. In a neural network approach, each function 
in the sample in turn is given by a
neural network which, when fed the values of the kinematic variables
as input, returns the value of the function itself as output.
The underlying idea is that neural networks can be used as
universal unbiased interpolators: starting from a Monte Carlo
representation of the probability density at the points where it is
known because there are data, they can be used to produce a
representation of the probability density everywhere in the data
region, and even to study the extrapolation outside it, and its breakdown.

In Refs.~\cite{f2ns,f2p} this strategy was tried off on a somewhat
simpler problem, namely, the construction of a parametrization of
existing data on the deep--inelastic structure function $F_2(x,Q^2)$
of the  proton and neutron.
In such case, one is only testing that the
method can be used to construct a faithful representation of the
probability density in a space of
functions, based on the measurement of the function at a finite
discrete number of points. The method was proven to be fast and
robust, to be amenable to detailed statistical studies, and to be in
many respects superior to conventional parametrizations of structure
functions based on a fixed functional form. However,
 the determination of a parton set involves 
the significant complication of having to go from one or more
physical observables to a set of parton
distributions. In order to achieve the result, one must tackle various
problems,  such
as  
deconvolution of hard coefficient
functions and resolution of evolution equations, and several further
subtleties such as the treatment of higher twist corrections and heavy
quark thresholds. All these issues are of course rather well
understood in the context of parton determinations, but their
implementation in a neural Monte Carlo framework is nontrivial, both
in principle and in practice.

In this paper, we present a first determination  of the nonsinglet
quark distribution from deep-inelastic data based on neural
networks. This result is the first step towards the determination of a
full parton distribution set. Furthermore, almost all of the
technical complications required for the construction of a neural
parton arise and have to be tackled already in the nonsinglet
case. Hence, this paper is a comprehensive introduction to 
neural parton fitting. 

The paper is organized as follows. In Sect.~2 we review the approach
of Refs.~\cite{f2ns,f2p} to the construction of a neural network
parametrization of a function based on its experimental sampling at a
finite set of points, we describe its application to the construction
of a parton set, thereby outlining the general strategy of the method,
and we summarize the main features of our approach and thus of our
results. In Sect.~3 we construct a Monte Carlo representation of the
information contained in the data, and test for its statistical
accuracy. In Sect.~4 we discuss a novel strategy to determine the
evolution of parton distributions and to combine them with
perturbative coefficients in order to determine the structure
functions, which combines the speed and reliability of Mellin
$N$-space evolution methods with the flexibility of $x$-space
evolution methods with respect to the choice of parton
parametrization. In Sect.~5 we turn to the neural networks which are
used to parametrize parton distributions, we discuss their features,
the training procedure whereby they are fitted to the data, and the
way this training is stopped before the noise in the data starts
affecting the results of the fit. In Sect.~6 we present our reference
next-to-leading order determination of the nonsinglet quark
distribution, we test for the accuracy of our estimate of central
value and statistical uncertainty, and we verify the stability of our
result upon the minimization strategy.  We discuss in particular how
we arrive at an accurate determination of the statistical properties
(specifically the uncertainty) of this determination. Potential
sources of theoretical uncertainty, in particular higher twists and
higher orders, are discussed in Sect.~7, while in Sect.~8 our
reference result is compared to results obtained at different
perturbative orders (leading and next-to-next-to-leading) and to
alternative existing determinations.

\section{General strategy}
\label{genstrat}

The method developed in Ref.~\cite{f2ns} provides a general framework
for the parametrization of physical observables by means of neural
networks. This method has since then been successfully applied to a diverse
variety of physical problems: structure functions~\cite{f2p}, spectral
functions for $\tau$ decays~\cite{tau}, energy spectra of B
decays~\cite{bdec}, and cosmic ray neutrino fluxes~\cite{neut}, thereby
proving its flexibility and robustness. Its application to parton
distributions is based on the same underlying concepts, but it is
significantly more intricate for a variety of reasons to be discussed
shortly, the most obvious being that parton distributions are not
directly physical observable quantities.

The general underlying
strategy of this approach 
is summarized in the flow-chart of Fig.~1, and, as in
Ref.~\cite{f2ns}, it involves
two distinct stages in order to go from the data to the
parton parametrization. In the first stage, a 
Monte Carlo sample of
replicas of the data  (``artificial data'') is generated. These can be
viewed as a sampling of the probability measure on the space of
physical observables (structure functions, cross sections, etc.) at
the discrete points where data exist. 
In the second stage one uses neural networks to interpolate between
points where data exist. When constructing a parton set, this
second stage in turn consists of two sub-steps: the determination of
physical observables from parton distributions (``evolution''), and
the comparison of the physical observables thus computed to the data
in order to  tune the best-fit form of input  neural parton distribution
(``training of the neural network''). Combining these two steps,  
the space of
physical observables   is mapped onto the space of parton
distributions, so the experimental information on the former can be
interpolated by neural networks in the latter.
Let us now describe each stage in turn, both in general and in the
specific case discussed in this paper.
\FIGURE[ht]{\epsfig{width=0.91\textwidth,
figure=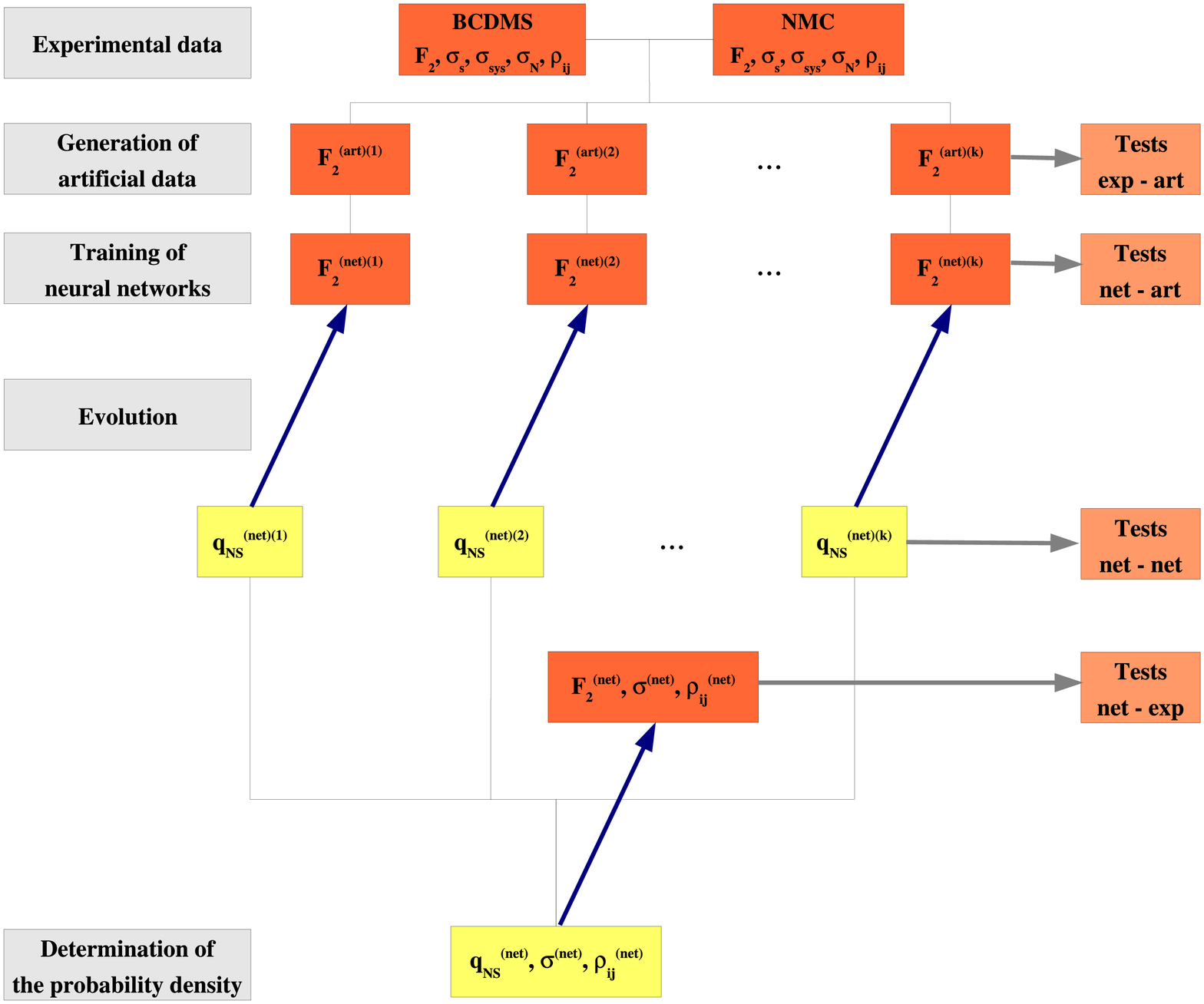} 
\caption{Flow-chart for the construction of a neural parton set.}
\label{summary}}


The starting experimental data in general consist of the
determination of several physical observables in distinct experiments,
each of which provides the measurement of the relevant quantity
at a discrete set of values of the kinematical variables. In general,
there will be a nontrivial set of correlations between determination
of different quantities: e.g., different observables determined at
different points may be correlated both between observables and
between points. In the current nonsinglet fit, we will consider a
single observable, namely the nonsinglet nucleon structure function 
defined as
\be
F_2^{\NS}(x,Q^2)\equiv F_2^p(x,Q^2)-F_2^d(x,Q^2) \ ,
\label{fnsdef}
\ee
in terms of the proton and deuteron structure functions.

The purpose of the artificial data generation is to produce
a Monte Carlo set of `pseudo--data', i.e.
$N_{\rep}$ replicas of the original set of
$N_{\dat}$ data points:
\be
F^{(\art)(k)}_i; \qquad k=1,\dots,N_{\rep},\quad  i=1,\dots,N_{\dat},
\label{replicas}
\ee where $F_i$ denotes one individual measurement: in our specific
case, a measurement of $F^{\NS}_2(x_i,Q^2_i)$.  The $N_{\rep}$ sets of
$N_{\dat}$ points are distributed according to an
$N_{\dat}$--dimensional multi-gaussian distribution about the original
points, with expectation values equal to the central experimental
values, and error and covariance equal to the corresponding
experimental quantities. Because the distribution of the experimental
data coincides (for a flat prior) with the probability distribution of
the value of the structure function at the points where it has been
measured~\cite{dagos,cowan}, this Monte Carlo set gives a sampling of
the probability measure at those points.  We can then generate
arbitrarily many sets of pseudo--data, and choose the number of sets
$N_{\rep}$ in such a way that the properties of the Monte Carlo sample
reproduce those of the original data set to arbitrary accuracy.  For
example, we can check that averages, variance and covariance of the
pseudo-data reproduce central values and covariance matrix elements of
the original data. This step is denoted by 'Tests exp-art' in
Fig.~1. Note that experimental errors (including correlated
systematics) must be treated as gaussian since this is what
experimental collaborations provide: the relevant issue is whether an
underlying non-gaussian distribution of best-fit parton distributions
can be obtained as a result of the fitting procedure. Whereas this
might be problematic if errors are determined as one--sigma contours
for a given parametrization, the Monte Carlo method gives considerably
more flexibility, as we now show.

The second step consists of 
training $N_{\rep}$  sets of neural networks. Each set contains all the
parton distributions which are being determined, as a function of $x$
at a given reference scale $Q_0^2$, and is based on all
the data in one single replica of the original data set. Thus, each
set will contain $N_{\pdf}$ parton distributions, determined from
$N_{\dat}$ data. The number of independent parton distributions is
$1\le N_{\pdf}\le 13$, and it
will depend on the specific data set which is used in the parton
determination. At most, there can be six  quark and
antiquark distributions  and
one gluon
distribution, but in practice not all of these may be accessible with
a given set of data.
Specifically, in the nonsinglet fit
discussed in this paper we will consider the simplest possible case,
in which each set contains only one parton
distribution, namely the C-even nonsinglet quark distribution
\be
q_{\NS}(x,Q_0^2)\equiv\left[u(x,Q_0^2)+\bar u(x,Q_0^2)-(d(x,Q_0^2)+\bar
  d(x,Q_0^2))\right] .
\label{qnsdef}
\ee

As in any parton fit, in order to compare the parton distributions
with the data, the PDFs must be evolved from the reference scale $Q_0^2$ to
the scale at which data are given, and combined with hard partonic
cross sections in order to obtain physical observables. Unlike in most
parton fits (such as e.g.~\cite{MRST,CTEQ,Alekhin}), however, each
parton distribution is parametrized by a neural network, rather than by a
fixed functional form. Then, the $N_{\dat}$ data in each replica are
used to determine the $N_{\pdf}$ neural networks of the corresponding
set. At the end of the procedure, we end up with $N_{\rep}$ sets of
$N_{\pdf}$ parton distributions, with each PDF given by a neural
network. The $N_{\rep}$ replicas of each parton distribution provide the
corresponding probability density: for example, the mean value of the
parton distribution at the starting scale for a given value of $x$ is
found by averaging over the replicas, and the uncertainty on this
value is the variance of the values given by the replicas.

The fit of the parton distributions to each replica of the data is
performed by maximum likelihood, by minimizing an error function, which
coincides with the $\chi^2$ of the experimental points when compared to
their theoretical determination obtained using the given set of parton
distributions. The  $\chi^2$ is computed by fully including the
covariance matrix of the correlated experimental uncertainties 
(with a suitable treatment of normalization errors~\cite{dagos}), as we
shall discuss in more detail in Sect.~5.1 below. Unlike in
conventional determinations of parton distributions with
errors~\cite{MRST,CTEQ,Alekhin}, 
however, the covariance matrices of the best--fit parameters are 
irrelevant and need not be computed. The uncertainty on the
final result is found from the variance of the Monte Carlo
sample. This eliminates the problem of choosing the value of
$\Delta\chi^2$ which corresponds to a one-sigma contour in the space
of parameters, a problem which becomes nontrivial in the presence of
incompatible data, underestimated errors, or unknown systematics, and which
therefore plagues current global fits~\cite{MRST,CTEQ}. 

Rather, one only has to make sure that each neural network provides a
consistent fit to its corresponding replica. If the underlying data
are incompatible or have underestimated errors, the best fit might be
worse than one would expect with properly estimated gaussian errors
--- for instance in the presence of underestimated errors it will have
typically a value of $\chi^2$ per degree of freedom larger than
one. If, on the contrary, experimental errors have been overestimated,
the final $\chi^2$ can turn out to be smaller than one per degree of
freedom.  Neural networks are ideally suited for providing a fit in
this situation, based on the reasonable assumption of smoothness: for
example, incompatible data or data with underestimated errors will
naturally be fitted less accurately by the neural
network~\cite{f2ns}. Also, this allows for non-gaussian behavior of
experimental uncertainties.  Indeed, whereas a neural network with
sufficiently large architecture can fit any compatible set of data,
the choice of a suitable stopping criterion ensures that only the
information contained in the data is fitted, but not the noise. This
can be done by verifying that the quality of the fit of data which
have been used in the fitting procedure is the same as the quality of
the fit to data which {\it have not} been used for fitting.

It is then possible to verify that each
neural network has the correct statistical features.
This step is denoted by
'Tests
net-art' in Fig.~1. Finally, the self-consistency of the Monte Carlo
sample can be tested in order to ascertain that it leads to consistent
estimates of the uncertainty on the final parton distributions, for
example by verifying that the value of the parton distribution
extracted from different replicas indeed behaves as a random variable
with the stated variance. This step is denoted by
'Tests
net-net' in Fig.~1. This set of tests allow us to make sure that the
Monte Carlo sample of neural networks provides a faithful and consistent
representation of the information contained in the data on the
probability measure in the space of parton distributions, and in
particular that the value of parton distributions and their
correlated uncertainties are correctly estimated.

Because of the use of neural networks for the parametrization of the parton
distributions, two technical aspects of our fitting procedure differ
from what is done in standard parton fits. The first is the method
which is used in order to evolve parton distributions by solving 
the QCD evolution equations. Indeed, we solve evolution equations in
Mellin moment space $N$~\cite{dflm}, since this allows for a fast and
accurate numerical solution. However, this method is usually applied
to parton parametrizations such that the Mellin transform of the
initial parton distribution can be computed analytically: a numerical
approach is not viable because in the region where the inverse Mellin
transform of the solution exists, the Mellin transform of the initial
condition does not converge. But a  neural network is a complicated
nonlinear function whose Mellin transform cannot be determined
analytically in closed form. The alternative of parametrizing directly
the initial parton distributions in $N$ moment space is unpalatable,
because one would need a parametrization in the complex plane of the
$N$ variable, whereas neural networks are most conveniently defined on a
compact space, such as $0\le x \le 1$.

 Therefore, we are led to introduce
a novel technique, which consists of determining the $x$--space
evolution kernel (Green function) from the $N$ space solution, in such
a way that any parton distribution can be evolved by convolution with
this universal kernel. This method combines the speed and reliability
of the $N$--space solution method, with the advantages of parametrizing
the parton distribution in $x$ space. 
Furthermore, because the kernel is universal
(i.e. independent of the specific boundary condition which is evolved)
it can be computed beforehand to arbitrarily high
accuracy and stored; the result can then be used during the fitting
procedure, thus leading to an extremely efficient evolution. This
evolution method will be discussed in detail in Sect.~4.

The other technical peculiarity which is motivated by the use of
neural networks is the choice of the minimization algorithm. Because
of the nonlinear dependence of the neural network on its parameters, and
the nonlocal dependence of the measured quantities on the neural network
(cross sections are expressed as convolution of the initial
distribution with evolution kernels and coefficient functions),
a genetic algorithm turns out to be the most efficient minimization
method. The use of a genetic algorithm is particularly convenient when
seeking a minimum in a very wide space with potentially many local
minima, because the method handles a population of solutions rather
than traversing a path in the space of solutions. The minimization
method will be discussed in detail in Sect.~5.

At the end of the minimization we end up with a Monte Carlo sample of
neural networks which provides our best representation of the measure in
the space of parton distribution. Generally, we
expect~\cite{f2ns,f2p} the uncertainty on 
our final result to be somewhat smaller than the uncertainty on the
input data, because the information contained in several data points is
combined. 
The properties of this measure can
be tested against the input data by using it to compute means,
variance and covariances which can be compared to the input
experimental ones which have been used in the parton
determination. Also, the 
stability of the result can be tested by excluding some
data points from the fit and checking that the best-fit result
predicts them correctly. 
These tests are denoted by
'Tests
net-exp' in Fig.~1, and they conclude our determination.

\section{Experimental data and Monte Carlo generation}
\label{data}
\subsection{Experimental data and kinematical cuts}
\label{datacuts}

The nonsinglet parton distribution $q_{\NS}(x,Q_0^2)$ 
can be extracted
from experimental data on proton and deuteron structure
functions $F_2^{p} (x,Q^2)$ and $F_2^{d} (x,Q^2)$. A detailed
discussion of all the available data was given 
in Refs.~\cite{f2ns,f2p}. In
short, early SLAC data have extremely large uncertainties, while data
from JLAB and the E665 collaboration~\cite{e665} are mostly at low
$Q^2$ and are almost entirely excluded by our kinematic cuts. This
leaves us with the data
published  
by the NMC \cite{nmc} and BCDMS
\cite{bcdms1,bcdms2} collaborations. We refer to~\cite{f2ns} for
a discussion of the kinematical and statistical features of these data,
in particular for the  details of their correlated systematics.

Once the systematics are known, the experimental covariance matrix for each
experiment can be easily computed
\be
\label{covmat}
{\rm cov}_{ij}=
\lp\sum_{p=1}^{N_{\sys}}\sigma_{i,p}\sigma_{j,p}+F_iF_j\sigma_N^2\rp+
\delta_{ij}\sigma_{i,s}^2 \ ,
\ee
where $F_i$, $F_j$ are 
central values, $\sigma_{i,p}$ are
the $N_{\sys}$ correlated systematic errors, $\sigma_N$ is the total
normalization uncertainty, and $\sigma_{i,s}$ is the statistical
uncertainty. The correlation matrix is  given by
\be
\rho_{ij}=\frac{ {\rm cov}_{ij}}{\sigma_{i,\tot}\sigma_{j,\tot}} \ ,
\label{cormat}
\ee
where the total error $\sigma_{i,\tot}$ for the $i$-th point is given by
\be
\sigma_{i,\tot}=\sqrt{\sigma_{i,s}^2+\sigma_{i,c}^2+ F_i^2\sigma_N^2} \ 
\label{toterr}
\ee
and the total correlated uncertainty $\sigma_{i,c}$ is the sum of all 
correlated systematics
\be
\sigma_{i,c}^2=\sum_{p=1}^{N_{\sys}} \sigma_{i,p}^2 \ .
\label{totsyst}
\ee

In order to keep higher twist corrections under control, only data
with $Q^2>$~3~GeV$^2$ are retained. No separate cut in the invariant
mass of the final state $W^2$ turns out to be necessary. A discussion
on the motivation for this choice of cuts and the possible effect of
their variation will be given in Sect.~6.

A scatter plot of the NMC and BCDMS  data in the $(x,Q^2)$ plane is
displayed in Fig.~\ref{kincov}, with points  excluded due to the $Q^2$
cut
marked with diamonds. Note that all these excluded data points
belong to the NMC experiment.
In Table \ref{datafeat}, the statistical features of the data (after
the cut)  are summarized.

\FIGURE[ht]{\epsfig{width=0.6\textwidth,figure=kincov.ps} 
        \caption{\small Kinematical coverage of the
available experimental data on the nonsinglet
structure function in the $(x,Q^2)$ plane.}
        \label{kincov}}

\TABLE[ht]{
\tiny
\begin{tabular}{|c|cc|c|ccccc|cc|} 
\hline
 Experiment & $x$ range & $Q^2$ range & $N_{\dat}$   
& $\la\sigma_{s}\ra$& $\la\sigma_{c}\ra$& 
$\la\frac{\sigma_{c}}{\sigma_{s}}\ra$ &
$\la \sigma_{N}\ra$   & $\la\sigma_{\tot}\ra$ 
&$\la\rho\ra$& $\la\mathrm{cov}\ra$ 
 \\
\hline
NMC & $9.0~10^{-3}~-~4.7~10^{-1}$ & $3.2-61$ GeV$^2$ &
 229 & 103.2 & 85.6  & 0.75  & 72.5  & 157.5  & 0.038  &  0.199
\\
\hline
BCDMS & $7.0~10^{-2}~-~7.5~10^{-1}$ & $8.7-230$ GeV$^2$ &
 254 & 22.3  & 12.2
& 0.50  & 22.3  & 37.3  & 0.163  &  0.085\\
\hline
\end{tabular}
\caption{\small Experimental data included in this analysis. All values of
  $\sigma$ and cov are given as percentages. Averages over all data
points  are 
noted with brackets $\langle \rangle$.}
\label{datafeat}
}

\TABLE[ht]{
\begin{tabular}{|c|ccc|} 
\multicolumn{4}{c}{
$F_2^{\NS}(x,Q^2)$}\\   
\hline
 $N_{\rep}$ & 10 & 100 & 1000 \\
\hline
$\la PE\lc\la F^{(\art)}\ra_{\rep}\rc\ra$ & 
20\% & 6.4\% & 1.3\%    \\
$r\lc F^{(\art)} \rc$ & 0.97 &  0.99& 0.99 \\
\hline
$\la V\lc \sigma^{(\art)}\rc\ra_{\dat}$ & $6.1~10^{-5}$ & $1.9~10^{-5}$
 & $6.7~10^{-6}$  \\
$\la PE\lc \sigma^{(\art)}\rc\ra_{\dat}$ & 33\%
& 11\% &  3\%   \\
$\la \sigma^{(\art)}\ra_{\dat}$ & 0.011 & 0.011 & 0.011 \\
$r\lc \sigma^{(\art)}\rc$ & 0.94 & 0.99  & 0.99 \\
\hline
$\la V\lc \rho^{(\art)}\rc\ra_{\dat}$ & $0.10$
&  $9.4~10^{-3}$ &  $1.0~10^{-3}$ \\
$\la \rho^{(\art)}\ra_{\dat}$ & 0.182 & 0.097 & 0.100\\
$r\lc \rho^{(\art)}\rc$ &  0.47&  0.79 & 0.97  \\
\hline
$\la V\lc \mathrm{cov}^{(\art)}\rc\ra_{\dat}$ & $ 5.5~10^{-9}$ 
&  $1.7~10^{-10}$  &  $5.7~10^{-11}$ \\
$\la \mathrm{cov}^{(\art)}\ra_{\dat}$ & 
$1.3~10^{-5}$ & $7.6~10^{-6}$ & $8.1~10^{-6}$ \\
$r\lc \mathrm{cov}^{(\art)}\rc$ & 0.41 &  0.81 & 0.98\\
\hline
\end{tabular}
\caption{\small 
Comparison between experimental and 
Monte Carlo data.\hfill\break
The experimental data have
$\la \sigma^{(\mrexp)}\ra_{\dat}=0.011$, $\la \rho^{(\mrexp)}\ra_{\dat}=
0.107$ and $\la \mathrm{cov}^{(\mrexp)}\ra_{\dat}=8.6~10^{-6}$. }
\label{gendata}
}

\subsection{Monte Carlo generation}
\label{mcgen}

The first step in our approach, as discussed in Sect.~\ref{genstrat},
is to generate a Monte Carlo sample of replicas of the experimental
data. Each Monte Carlo replica of the original experimental data is
generated as a multi-gaussian distribution. More precisely, for
each data point $F_i^{\rm (exp)}\equiv F_2^{\NS}(x_i,Q^2_i)$
we generate $k=1,\ldots,N_\rep$  artificial points $F^{(\art)(k)}$
as follows
\be
\label{freplicas}
F_i^{(\art)(k)}=\lp1+r_{N}^{(k)}\sigma_N\rp\lp F_i^{\rm (\mrexp)}+
 \sum_{p=1}^{N_{\sys}}r_{p}^{(k)}\sigma_{i,p}+r_{i}^{(k)}\sigma_{i,s}\rp
 \ , \ k=1,\ldots,N_{\rep} \ ,
\ee
using independent univariate gaussian random numbers
$r^{(k)}$ for each independent error source.

The value of $N_{\rep}$ is determined in such a way that the Monte
Carlo set of replicas models faithfully the probability distribution
of the data in the original set. A quantitative comparison can be
performed by defining suitable statistical estimators (see
Appendix \ref{dataest}). Results are presented in Table~\ref{gendata}. 
The results
show that a sample of 1000 replicas is sufficient to ensure average
scatter correlations of 99\% and accuracies of a few percent on
structure functions, errors and correlations.

Normalization errors are not included in the covariance matrix on the
same footing as other sources of systematics since,
as by now  well known~\cite{dagos}, this would bias the fit.
Rather, normalization errors are included by rescaling 
all errors (statistical $\sigma_{i,s}$ 
and each systematic  $\sigma_{i,p}$)  independently for each replica
by a factor
\bea
\overline \sigma_{i,s}^{(k)}&=& (1+r^{(k)}_N\sigma_N)\sigma_{i,s} \ ,
\nonumber\\
\overline \sigma_{i,p}^{(k)}&=& (1+r^{(k)}_N\sigma_N)\sigma_{i,p}
\qquad p=1 \ ,\ldots,N_\sys \ ,
\eea
where $r^{(k)}_N$ is the random variable Eq.~(\ref{freplicas}) associated to
the normalization uncertainty. The covariance matrix is then given by
\be
\label{covmatnn}
{\overline\mathrm{cov}^{(k)}}_{ij}=
\lp\sum_{p=1}^{N_{\sys}}{\overline\sigma^{(k)}}_{i,p}
{\overline\sigma^{(k)}}_{j,p}\rp+\delta_{ij}{\overline\sigma^{(k)2}}_{i,s} \ ,
\ee
in terms of the rescaled uncertainties.

\section{From parton distributions to physical observables}
\label{evolution}
The nonsinglet structure function $F_2^{\NS}(x,Q^2)$ Eq.~(\ref{fnsdef}) 
is determined from
 the nonsinglet quark distribution $q_{\NS}(x,Q_0^2)$ Eq.~(\ref{qnsdef}) 
at a reference
 scale $Q_0^2$ by first evolving the parton distribution to the scale
 $Q^2$ by means of QCD evolution equations, and then convoluting it
 with the appropriate hard coefficient function $C_{\NS}(x,\aq)$:
\be
\label{coef}
F_2^{\NS}(x,Q^2)=\frac{1}{6}
x\int_x^1 \frac{dy}{y}C_{\NS}(y,\aq)q_{\NS}\lp
\frac{x}{y},Q^2\rp \ .
\ee
Both coefficient functions and the anomalous dimensions which drive
perturbative evolution are now known up to next-to-next-to leading
order (and in fact the coefficient function up to NNNLO~\cite{nnnlo}).
 Therefore, we will provide determinations of the parton
distributions at leading (LO), next-to-leading (NLO) and   next-to-next-to
 leading
order (NNLO). In this section we will discuss in detail the methods
that we use in order to actually determine the structure function from
the input parton distribution, specifically the resolution of
evolution equations. Only the nonsinglet case will be discussed
explicitly, while the (straightforward) generalization to the singlet
case will be discussed in a forthcoming publication.

\subsection{Leading-twist evolution and coefficient functions}
\label{ltev}

As well known, QCD evolution equations are most easily written for
Mellin moments of the parton distributions, defined as
\be
\label{meldef}
q_{\NS}(N,Q^2)\equiv\int_0^1 \!dx\,x^{N-1} q_{\NS} (x,Q^2) \ ,
\ee
where by slight abuse of notation we denote the function and its
transform with the same symbol.
They take the form\be
\label{evolN}
\frac{d}{d\ln Q^2}q_{\NS}(N,Q^2)=\frac{\aq}{4\pi}\gamma_{\NS}
(N,\aq)\, q_{\NS}(N,Q^2) \ ,
\ee
where the anomalous dimension can be expanded in powers of $\aq$
as
\be
\label{andim}
\gamma_{\NS}(N,\aq)=\gamma^{(0)}_{\NS}(N)+\frac{\aq}{4\pi}\gamma^{(1)}_{\NS}(N)
+\lp \frac{\aq}{4\pi}\rp^2\gamma^{(2)}_{\NS}(N)+\dots
\ee
The NNLO contribution has been computed recently in
Ref. \cite{mvvns}.
Defining analogously the Mellin transforms of coefficient function and
structure function, the latter is given by
\be
\label{f2co}
F_2^{\NS}(N-1,Q^2)=\frac{1}{6}
C_{\NS}(N,\aq)q_{\NS}\lp N,Q^2\rp ,
\ee
with the perturbative expansion of the coefficient function
\be
 C_{\NS}(N,\aq)=1+\frac{\aq}{4\pi}C_{\NS}^{(1)}(N)+
\lp \frac{\aq}{4\pi}\rp^2 C_{\NS}^{(2)}(N)+\dots .
\ee

It appears therefore especially convenient~\cite{dflm, pegasus} to determine
the structure function in Mellin space and then invert the Mellin
transform, since the $N$ space expression of the structure function
can be determined in closed
form, and the problem is reduced to the computation of the single
Mellin inversion integral. In $x$-space instead one has to solve an
integro--differential equation. Even though efficient numerical
methods have been developed for the resolution of the
integro-differential form of the evolution equation 
(see e.g.~\cite{qcdnum,hoppet}), the only
genuine advantage of an $x$--space approach is the possibility to
work directly with an  $x$--space parametrization of parton
distributions. As discussed in Sect.~2, this option is especially
convenient when using neural networks, since these are most easily defined as
function on a compact space such as $0\le x\le 1$.

However, we observe that the solution to the evolution
equation~(\ref{evolN}) has the form
\be
\label{evoln}
q_{\NS}(N,Q^2)=\Gamma(N,\aq,\aqq)q_{\NS}(N,Q^2_0) \ ,
\ee
where the evolution kernel $\Gamma(N,\aq,\aqq)$ is entirely
determined by the anomalous dimension $\gamma_{\NS}(N,\aq)$, and it does
not depend on the boundary condition $q_{\NS}(N,Q^2_0)$. Hence, if we
take an inverse Mellin transform of Eq.~(\ref{evoln}) we get
\be
\label{evolx}
q_{\NS}(x,Q^2)=
\int_x^1 \frac{dy}{y}\Gamma(y,\aq,\aqq)q_{\NS}\lp \frac{x}{y},Q_0^2\rp .
\ee
We can then determine the kernel $\Gamma(x,\aq,\aqq)$ explicitly by
Mellin inversion, and use it to evolve any $x$--space parton
distribution using Eq.~(\ref{evolx}). The only complication is related
to the large-$x$ behavior of 
the kernel $\Gamma(x,\aq,\aqq)$.

Furthermore, we can express the structure function in terms of the
initial quark distribution as
\be
\label{f2cogam}
F_2^{\NS}(N-1,Q^2)=\frac{1}{6} \widetilde\Gamma(N,\aq,\aqq)\,q_{\NS}(N,Q^2_0) \ ,
\ee
where
\be
\label{gamtildef}
\widetilde\Gamma(N,\aq,\aqq)\equiv C_{\NS}(N,\aq)\Gamma(N,\aq,\aqq),
\ee
i.e. absorbing the coefficient function in a redefined evolution
kernel. Again, we can determine the Mellin inverse of this redefined
kernel, and use it to obtain the measured structure function from the
initial $x$--space parton distribution as 
\be
\label{f2conv}
F_2^{\NS}(x,Q^2)=\frac{1}{6} x\int_x^1 \frac{dy}{y}
\tilde{\Gamma}(y,\aq,\aqq)\,q_{\NS}\lp \frac{x}{y},Q_0^2\rp \ .
\ee

The evolution kernels can be computed, interpolated to arbitrary
accuracy and stored, and then used to evolve the initial PDF or
determine the structure function from it through the computation of a
single real convolution integral, eqs.~(\ref{evolx}) and 
(\ref{f2conv}).

\subsection{Solution of the evolution equation in $N$ space}
\label{evn}

We determine the evolution kernel up to NNLO in terms of the running
coupling, for which we take the expanded solution of the
renormalization group equation, i.e., up to NNLO
\bea
\aq&=&\aq_{LO}\Bigg[ 1+\aq_{LO}\left[ \aq_{LO}- \alpha_s\lp M_Z^2\rp \right]
(b_2-b_1^2) \nonumber \\
&+&\aq_{NLO}b_1 \ln\frac{\aq_{NLO}}{\alpha_s\lp M_Z^2\rp}\Bigg],
\eea
with 
\be
\aq_{NLO}=\aq_{LO}\left[ 1-b_1\aq_{LO}\ln\lp 1+\beta_0\alpha_s\lp M_Z^2\rp
\ln\frac{Q^2}{M_Z^2}\rp\right] \ ,
\ee
\be
\aq_{LO}=\frac{\alpha_s\lp M_Z^2\rp}{ 1+\beta_0\alpha_s\lp M_Z^2\rp
\ln\frac{Q^2}{M_Z^2}}, 
\ee
and the beta function coefficients given by 
\be
Q^2\frac{da_s(Q^2)}{d Q^2}=-\sum_{k=0}^2 \beta_k a_s(Q^2)^{k+2} , \qquad
a_s(Q^2)=\frac{\alpha_s(Q^2)}{4\pi} \,,
\ee
where
\bea
\beta_0 &=& 11-\frac{2}{3}N_f \ , \\
\beta_1 &=& 102-\frac{38}{3}N_f \ ,\\
\beta_2 &=& \frac{2857}{2}-\frac{5033}{18}N_f+\frac{325}{54}N_f^2 \ , 
\eea
and $b_i\equiv \beta_i/\beta_0$. We will use the current PDG~\cite{pdg}
average value of $\aq$, namely
\be
\label{aq}
\alpha_s(M_Z^2)=0.118 \pm 0.002
\ee
with $M_Z=91.187$ GeV.

The evolution factor is then explicitly given by
\bea
\label{gammaN}
&&\Gamma(N,\aq,\aqq)=\lp \frac{\aq}{\aqq}\rp^{-\gamma^{(0)}(N)/\beta_0}
\Bigg( 1 -\lp a_s(Q^2)-  a_s(Q^2_0)\rp U_{\NS}^{(1)}(N)
\nonumber
\\
&&\qquad+\lp a^2_s(Q^2)-  a^2_s(Q^2_0)\rp U_{\NS}^{(2)}(N)
-a_s(Q^2)a_s(Q_0^2) \lp U_{\NS}^{(1)}(N)\rp^2
\Bigg) \ ,
\eea
where (adopting the notations and conventions of Ref. \cite{pegasus})
we have  introduced
nonsinglet evolution coefficients 
\be
U_{\NS}^{(1)}(N)=-\frac{1}{\beta_0}\lp \gamma^{(1)}(N)-
b_1\gamma^{(0)}(N)\rp \ ,
\ee
\be
U_{\NS}^{(2)}(N)= -\frac{1}{2}\left[ \frac{1}{\beta_0}\gamma^{(2)}(N)
-b_1U_{\NS}^{(1)}-\frac{b_2}{\beta_0}\gamma^{(0)}(N) 
- \left(U_{\NS}^{(1)}(N)\right)^2\right] \,.
\ee

At NNLO, both the strong coupling and the
parton distributions are discontinuous when crossing 
heavy quark thresholds. 
We set the thresholds 
$m_c=1.4 ~{\rm GeV}$, $m_b=4.5 ~{\rm GeV}$ and
$m_t=175 ~{\rm GeV}$ with, at NNLO,
the discontinuity
\be
\alpha_{s,f+1}(m_{f}^2)=\alpha_{s,f}(m_{f}^2)+\lp \frac{C_2}{4\pi} \rp^2
\alpha_{s,f}(m_{f}^2)^3 \ ,
\ee
where $C_2=14/3$ \cite{coefalphannlo}. We match parton distributions
at threshold as
\be
q_{\NS}^{(n_f+1)}(N,m_h^2)=q_{\NS}^{(n_f)}(N,m_h^2) \ .
\ee
Heavy quark mass terms are included in the hard coefficients, and in
the nonsinglet sector they vanish up to NLO, while at NNLO they are
not known (though they can be approximated from their large $x$
behavior~\cite{thornehq}). Therefore, 
 we defer the inclusion of heavy quark mass effects to a forthcoming
 full fit including the singlet contribution, hence, for all practical
 purposes, we use a zero--mass variable flavour number scheme in the
 nonsinglet sector. Note that because of
 this, the nonsinglet structure function $F_2^{\NS}$ has a NNLO
 discontinuity at the heavy quark thresholds.

\subsection{Evolution kernel in $x$-space}
\label{evkx}

\FIGURE[ht]{\epsfig{width=0.58\textwidth,
figure=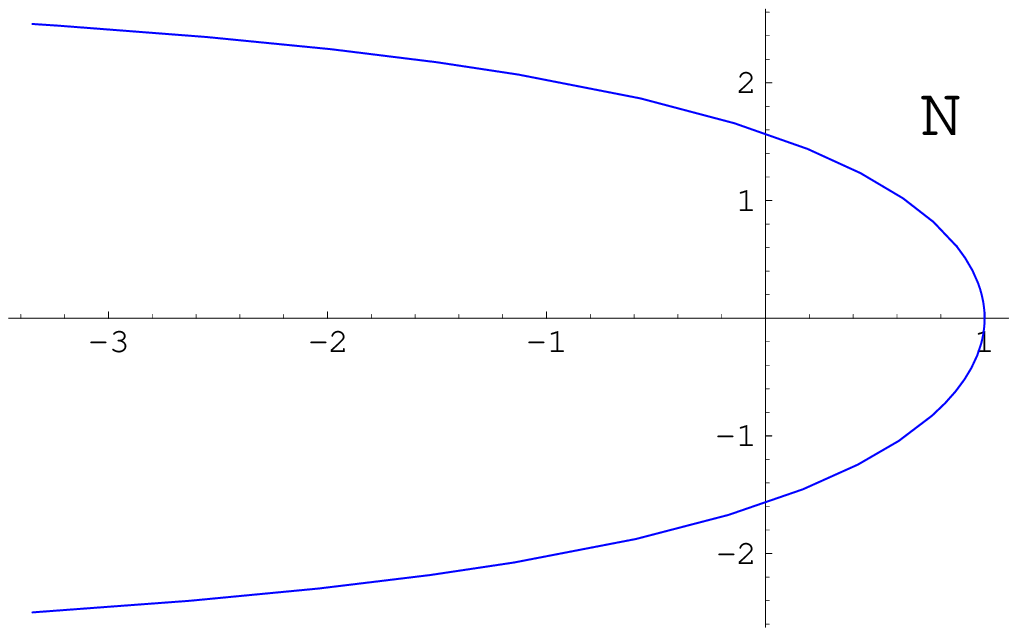} 
\caption{The path in the complex N-space followed by the
Talbot integration path, Eq.~(\ref{tbpath}) for $r=1$.}
\label{talbotpath}}

The $x$--space evolution factor is determined by numerical
computation of the Mellin inverse
\be
\label{gammaX}
\Gamma(x,\aq,\aqq)=\int_{c-i\infty}^{c+i\infty}
\frac{dN}{2\pi i}x^{-N}\Gamma(N,\aq,\aqq) \ ,
\ee
of the $N$ space evolution factor $\Gamma(N,\aq,\aqq)$,
Eq.~(\ref{gammaN}), as well as of the associate $\tilde \Gamma(N,\aq,\aqq)$,
Eq.~(\ref{gamtildef}).
The numerical computation of the Mellin inversion integral is 
delicate, because the oscillatory behavior of the integrand at large
$x$ is not damped by multiplication by an initial PDF.
This problem is mitigated by a suitable choice of integration path.
We choose the Talbot path,
defined by the condition 
\be
\label{tbpath}
N(\theta)=r\theta \lp 1/\tan \theta +i\rp, \quad  -\pi \le \theta \le \pi \ ,
\ee
with $r$ a constant, shown in Fig.~\ref{talbotpath}.

To further improve the numerical efficiency the Fixed Talbot
algorithm can be used \cite{ft} where the integral
is replaced by the sum
\be
\Gamma\lp x\rp=\frac{r}{M}
\lc \frac{1}{2}\Gamma\lp N=r\rp x^{-r}+
\sum_{k=1}^{M-1}  \mathrm{Re}
\lc x^{-N\lp \theta_k\rp} 
\Gamma\lp N\lp \theta_k\rp\rp 
\lp 1+i\sigma\lp \theta_k\rp\rp\rc\rc
\ee
where 
$\sigma\lp \theta\rp\equiv \theta+\lp \theta/\tan\theta
-1\rp/\tan\theta$,
and $\theta_k=k\pi/M$, and
 $r=2M/\lp 5\ln 1/x \rp$. It can be shown that $M$ is the relative
accuracy, i.e. the number of
accurate digits. We shall take $M=16$.
The $x-$space evolution kernel
$\Gamma(x)$ determined thus is displayed in Fig.~\ref{gammaxplots},
at different perturbative orders and in
Fig.~\ref{gammaplots2} for different values of $Q^2$.
\DOUBLEFIGURE[ht]{evol1.ps,width=0.48\textwidth}
{evol2.ps,width=0.48\textwidth}
       {\label{gammaxplots} Non-singlet evolution kernel $\Gamma(x)$
         computed at 
different perturbative orders in the
kinematical region relevant to nonsinglet evolution.
The evolution scales are $Q_0^2=2$ GeV$^2$ and
$Q^2=10^4$ GeV$^2$.}{\label{gammaplots2}
Same as Fig.~\ref{gammaxplots} for the LO
dependence of the evolution factor on the evolution
length $Q^2$. As in the previous case, $Q_0^2=2$ GeV$^2$. }

The determination of the evolution kernel presented so far
only holds for $x<1$.
As well known (see Appendix A), the Mellin inverse of the
anomalous dimension Eq.~(\ref{andim}) only exists for $x<1$, while at
$x=1$ it behaves as a distribution. Because the $N$--space 
evolution kernel is
obtained by exponentiating the anomalous dimension, its Mellin
transform must be defined through its action on a test function. In
particular, we must check that 
integration over the kernel converges as $x\to 1$.
To this purpose, we define a subtracted kernel $\Gamma^{(+)}(x,\aq,\aqq)$
\bea\label{gamgp}
\Gamma(x,\aq,\aqq)&=&\Gamma^{(+)}(x,\aq,\aqq)+G \delta(1-x) \ , \\
\label{gpdef}
\Gamma^{(+)}(x,\aq,\aqq)&\equiv&\Gamma(x,\aq,\aqq)-G \delta(1-x).
\eea
The constant $G$ is defined as
\be\label{gdef}
G \equiv\int_0^1 dx  \Gamma(x,\aq,\aqq) = \Gamma(N,\aq,\aqq)\Big|_{N=1}\ ,
\ee
where $\Gamma(N,\aq,\aqq)$ is given by Eq.~(\ref{gammaN}), and it
converges at $N=1$.

The evolution equation, Eq.(~\ref{evolx}), in terms of the subtracted kernel
takes the form 
\bea
\label{evolfull}
q(x,Q^2)&=& G q(x,Q_0^2) + \int_x^1
\frac{dy}{y}\Gamma^{(+)}(y,\aq,\aqq)q\lp
\frac{x}{y},Q_0^2\rp\nonumber\\ 
&=&
q(x,Q_0^2)\lp  G-\int_0^x dy \Gamma(y,\aq,\aqq)\rp \nonumber \\
&+& \int_x^1 \frac{dy}{y}
\Gamma(y,\aq,\aqq)\lp q\lp \frac{x}{y},Q_0^2\rp- y q(x,Q_0^2)\rp. 
\eea
It is easy to prove  that all integrals in
Eq.~(\ref{evolfull}) converge and can be computed numerically,
exploiting the fact that the behavior of
the evolution kernel in the large $x$ limit is known from soft
gluon resummation arguments. An explicit proof is given in Appendix A.

Hence we will use Eq.~(\ref{evolfull}) to determine the evolution of
the quark distribution. The numerical accuracy of this method is 
tested by comparing it to  the
benchmark evolution tables first presented in Ref.~\cite{lh2} and
recently updated including the full NNLO anomalous dimension in
Ref.~\cite{heralhc}. In order to allow for a direct comparison to the
benchmark tables, we
modify some of our final choices in the solution of the evolution
equations. Specifically, we fix the number of flavors (to $n_f=4$)
and we keep the NLO contributions to the evolution factor
Eq.~(\ref{gammaN}) in unexpanded form. With these choices, we 
determine the evolution
of the  $u$ valence
parton distribution.  The results and the accuracy of this benchmark
are shown in Table~\ref{lh}, where we use the same
parameters as  in \cite{lh2,heralhc}.  More details about the
benchmarking of QCD evolution codes can be found in 
Ref.~\cite{heralhc}. We find an accuracy of order
$\mathcal{O}\lp 10^{-5} \rp$, comparable to that of current
evolution codes. The same result has been obtained for the
$d$ valence distribution. Comparable accuracy is expected in $C$--even
evolution, relevant for our paper (not included in the benchmark~\cite{lh2})

\TABLE[ht]{
\begin{tabular}{|c|c|cc|} 
\hline
x & $xu_v(x,Q^2_0)$ (LH) &  $xu_v(x,Q^2_0)$ (NNPDF) & Rel. error\\
\hline
\multicolumn{4}{c}{Leading order}\\
\hline
$10^{-7}$   &   $5.7722~10^{-5}$    & $5.7722~10^{-5}$  & $3.3760~10^{-6}$  \\
$10^{-6}$   &   $3.3373~10^{-4}$    & $3.3373~10^{-4}$&$1.6880~10^{-6}$    \\
$10^{-5}$   &   $1.8724~10^{-3}$    & $1.8724~10^{-3}$&$1.9212~10^{-6}$ \\
$10^{-4}$   &   $1.0057~10^{-2}$    & $1.0057~10^{-2}$&$   
1.4095~10^{-6}$    \\
$10^{-3}$   &   $5.0392~10^{-2}$    & $5.0392~10^{-2} $&$  2.6145~10^{-6} $  
   \\
$10^{-2}$   &   $2.1955~10^{-1}$    &  $2.1955~10^{-1} $&$3.1065~10^{-6} $   
  \\
$0.1$       &   $5.7267~10^{-1}$    & $5.7267~10^{-1} $&$  6.4524~10^{-6}   $ 
 \\
$0.3$       &   $3.7925~10^{-1}$    & $3.7925~10^{-1}$&$   9.2674~10^{-6}    $
   \\
$0.5$       &   $1.3476~10^{-1}$    &$1.3476~10^{-1} $&$  1.1307~10^{-5}$\\
$0.7$       &   $2.3123~10^{-2}$    & $2.3122~10^{-2} $&$  2.1165~10^{-5}$    
 \\
$0.9$       &   $4.3443~10^{-4}$ &$4.3440~10^{-4} $&$6.3630~10^{-5}$       \\
\hline
\multicolumn{4}{c}{Next-to-Leading order}\\
\hline
$10^{-7}$   & $1.0616~10^{-4}$  &  $1.0616~10^{-4}$&$   2.1462~10^{-6}$    \\
$10^{-6}$   & $5.4177~10^{-4}$ &  $5.4177~10^{-4} $&$  8.7799~10^{-6}$      \\
$10^{-5}$   & $2.6870~10^{-3}$ &  $2.6870~10^{-3}$&$   9.7796~10^{-6}$      \\
$10^{-4}$   & $1.2841~10^{-2}$ &  $1.2841~10^{-2}$&$   1.3380~10^{-5}$      \\
$10^{-3}$   & $5.7926~10^{-2}$&  $5.7926~10^{-2}$&$   8.5063~10^{-6}$        \\
$10^{-2}$   & $2.3026~10^{-1}$&  $2.3026~10^{-1}$&$   3.0757~10^{-7}$       \\
$0.1$       & $5.5452~10^{-1}$&  $5.5452~10^{-1}$&$   7.6419~10^{-7}$     \\
$0.3$       & $3.5393~10^{-1}$&  $3.5393~10^{-1}$&$   2.6979~10^{-6}$        \\
$0.5$       & $1.2271~10^{-1}$&$1.2271~10^{-1}$&$   2.4466~10^{-5}$        \\
$0.7$       & $2.0429~10^{-2}$ &  $2.0429~10^{-2}$&$   1.4810~10^{-5}$       \\
$0.9$       & $3.6096~10^{-4}$ &  $3.6094~10^{-4}$&$   6.0762~10^{-5}$       \\
\hline
\multicolumn{4}{c}{Next-to-Next-to-Leading order}      \\
\hline
$10^{-7}$   & $1.5287~10^{-4}$&   $1.5287~10^{-4}$&$   1.5497~10^{-5}$    \\
$10^{-6}$   & $6.9176~10^{-4}$  & $6.9176~10^{-4}$&$   5.0711~10^{-6}$    \\
$10^{-5}$   & $3.0981~10^{-3}$  & $3.0981~10^{-3}$&$   9.5455~10^{-6}$    \\
$10^{-4}$   & $1.3722~10^{-2}$  & $1.3722~10^{-2}$&$   1.8022~10^{-5}$   \\
$10^{-3}$   & $5.9160~10^{-2}$ & $5.9160~10^{-2}$&$   5.0631~10^{-6}$      \\
$10^{-2}$   & $2.3078~10^{-1}$  & $2.3078~10^{-1}$&$   2.4853~10^{-6}$    \\
$0.1$       & $5.5177~10^{-1}$   & $5.5177~10^{-1}$&$   2.4747~10^{-6}$    \\
$0.3$       & $3.5071~10^{-1}$  & $3.5071~10^{-1}$&$   2.8430~10^{-7}$     \\
$0.5$       & $1.2117~10^{-1}$   & $1.2117~10^{-1}$&$   3.5893~10^{-5}$    \\
$0.7$       & $2.0077~10^{-2}$  & $2.0077~10^{-2}$&$   5.5823~10^{-6}$    \\
$0.9$       & $3.5111~10^{-4}$  & $3.5109~10^{-4}$&$   5.8172~10^{-5}$    \\
\hline
\end{tabular}
\caption{Comparison of the Les Houches parton evolution benchmark
results (LH) with our results (NNPDF).}
\label{lh}
}

The solution of the evolution equation
through the determination of an evolution factor is particularly
efficient because of the universality of the evolution factor itself,
i.e., its independence of the specific boundary condition which is
being evolved. Hence, the evolution factor can be precomputed and
stored, and then used during the process of parton fitting or when
evolving different parton distributions, without having to recompute
it each time.

This fact can be exploited in an optimal way during parton
fitting and parton evolution. In parton fitting, a given boundary
condition must be evolved many times up to the fixed pairs of values
of $(x,Q^2)$ at which data are available (in our case, those shown in
Fig.~\ref{kincov}). Namely, for the $i$-th data point, one must
compute $q(x_i,Q^2_i)$ given by Eq.~(\ref{evolfull}). The
constant $G$ Eq.~(\ref{gdef}) (which depends only on $Q^2_i$) can be
precomputed and stored for all required values of $Q^2$. Furthermore,
the numerical determination of the 
integral over $y$ on the right--hand side of Eq.~(\ref{evolfull}) involves the
determination of the integrand at a set of values of $y=y_{k i}$, which in
turn depend on the given values of $x_i,Q^2_i$:
$y_{ki}=y_{ki}(x_i,Q^2_i)$. 
The integrand is fully determined by knowledge of 
the following values of initial parton distribution and of the
evolution kernel:
\bea
\label{gamikdef}
\Gamma_{ik}&\equiv& \Gamma(y_{ik},\alpha_s(Q_i^2),\aqq)\\
\label{qikdef}
\tilde q_{ik}&\equiv& q \lp \frac{x_i}{y_{ik}},Q_0^2\rp-
 y_{ik} q(x_i,Q_0^2).
\eea

The most computing--intensive task is the determination of the 
values $\Gamma_{ik}$ Eq.~(\ref{gamikdef}) of  evolution kernel. We have
precomputed and stored these values for all the necessary values of
$x_i$, $Q^2_i$ and  $y_{ik}$. In
practice, we use gaussian integration with $N_{\rm pt}=
16\left(2^{N_{\rm it}+1}-1\right)$ points and determine the values of
$y$ accordingly. In
Table~\ref{evintacctable}  we show the percentage deviation $\epsilon$
of the left-hand side of Eq.~(\ref{evolfull}), when
compared to what is obtained from the use of a library routine with nominal 
percentage accuracy of $10^{-5}$. It is apparent that $N_{\rm it}=4$,
i.e. integration with about 500 points is sufficient to reach this accuracy.

\TABLE{
\begin{tabular}{|ccccc|}
\hline
$N_{\rm it}$ &  $\la \epsilon \ra$ & $\sigma_{\epsilon}$
& $\epsilon_{\rm max}$ & \# of points with $\epsilon \ge 5~10^{-5}$. \\
\hline
1  &  $2.4~10^{-2}$ & $8.5~10^{-2}$& $6.6~10^{-1}$ & 21.3\%\\
2  & $1.3~10^{-4}$ & $6.2~10^{-4}$ & $4.0~10^{-3}$ & 7.2\% \\
4  & $1.6~10^{-5}$ & $1.5~10^{-4}$ & $3.3~10^{-3}$ &5.4\% \\
6  & $1.6~10^{-5}$ &  $1.5~10^{-4}$ & $3.3~10^{-3}$  & 5.4\% \\
\hline
\end{tabular}
\caption{Relative accuracy $\epsilon$ (see text) in the result obtained for
NLO evolution as a function of the number of points $N_{\rm pt}=
16\left(2^{N_{\rm it}+1}-1\right)$
used for gaussian integration.}
\label{evintacctable}
}

This method is of course only convenient when evolving up to a fixed
set of points in the $(x,Q^2)$ plane. When evolving up a parton
distribution, e.g. in order to input it to some other computation, it
is instead convenient to precompute a full interpolation of the kernel
$\Gamma(x,\aq,\aqq)$. Evolution is then reduced to the determination
of the convolution Eq.~(\ref{evolfull}) of this kernel with the parton
distribution. This is somewhat less efficient than the previous method
because it requires the computation of this convolution over the
interpolation each time a point is evolved, but it already avoids the
main bottleneck, namely, the Mellin inversion of the kernel. We have
interpolated the kernel $\Gamma(x,\aq,\aqq)$ and its integral up to
$x$ which appears in the first term of Eq.~(\ref{evolfull}), using
Chebyshev polynomials. 

In order to speed up  the interpolation, the leading large $x$ 
behavior determined at each perturbative order
 (as described in Appendix A) is
divided out before interpolating, which considerably smoothes the
large $x$ growth of $\Gamma(x,\aq,\aqq)$.
We have found that the use of 200 polynomials
for the interpolation considered above produces results with
a precision of
 $\mathcal{O}\lp 10^{-5}\rp$,
for all the $(x,Q^2)$ range covered by experimental data. This accuracy
is enough for present purposes and it could be improved by increasing
the number of polynomials.  In Table \ref{f2intertable}
we compare results obtained with the interpolated
evolution factors to the exact evolution for
different values of $x$, with the input parton distribution set equal to our
best-fit nonsinglet neural parton distribution, to be described in
Sect.~6.

\TABLE{
\begin{tabular}{|c|c|cc|} 
\hline
x & $F_2^{\NS}(x,Q^2)$ (Exact) &  $F_2^{\NS}(x,Q^2))$ 
(Interpolated) & Rel. error\\
\hline
\multicolumn{4}{c}{Next-to-Next-to-Leading order}      \\
\hline
$10^{-3}$   & $3.7288~10^{-2}$& $3.7288~10^{-2}$& 
$1~10^{-6}$       \\
$10^{-2}$   & $1.1750~10^{-2}$& $1.1750~10^{-2}$& 
$7~10^{-6}$             \\
$0.1$       & $3.3497~10^{-2}$& $3.3497~10^{-2}$& 
$1~10^{-5}$        \\
$0.3$       & $4.4425~10^{-2}$& $4.4425~10^{-2}$& 
$3~10^{-7}$            \\
$0.5$        & $2.2159~10^{-2}$& $2.2159~10^{-2}$& 
$3~10^{-5}$         \\
$0.7$         & $4.7996~10^{-3}$& $4.7996~10^{-2}$& 
$6~10^{-5}$         \\
$0.9$        & $1.5031~10^{-4}$& $1.5030~10^{-4}$& 
$9~10^{-5}$          \\
\hline
\end{tabular}
\caption{Comparison of the results for exact evolution
at NNLO with coefficient function to
 to results obtained interpolated evolution (see text). 
Evolution is performed from
$Q_0^2=2~ \mathrm{GeV}^2$ to $Q^2=230~ \mathrm{GeV}^2$. The input is
taken to be equal to the final central fit discussed
in
Sect.~6.}
\label{f2intertable}
}

We shall choose $Q_0^2=2 ~{\rm GeV}^2$ as a starting scale for
perturbative evolution. 
We have explicitly checked the independence of 
the results with respect this choice.

\subsection{Target  mass corrections and higher twists}
\label{tmcht}
Even though we discard data at very low $Q^2$ data in order to
minimize the impact of higher twist corrections, target-mass
corrections~\cite{tmc}, which are the dominant higher twist
corrections and are known in closed form should be included in order to
increase the accuracy in the low--$Q^2$ region. The twist-four
(i.e. next-to-leading twist) structure function with the inclusion of
target-mass corrections is given by
\be
F_2^{\rm NLT}(x,Q^2)=
\frac{x^2}{\tau^{3/2}}\frac{F_2^{\rm LT}(\xi_{\rm
    TMC},Q^2)}{\xi^2_{\rm TMC}}
+6\frac{M^2}{Q^2}\frac{x^3}{\tau^2}I_2(\xi_{\rm TMC},Q^2) \ ,
\label{tmc}
\ee
where 
\be
\label{i2}
I_2(\xi_{\rm TMC},Q^2)=\int_{\xi_{\rm TMC}}^1
dz\frac{F_2^{\rm LT}(z,Q^2)}{z^2} \ ,
\ee
\be
\xi_{\rm TMC}=\frac{2x}{1+\sqrt{\tau}},\qquad \tau=1+\frac{4M^2x^2}{Q^2},
\ee
where $F_2^{LT}(x,Q^2)$ is the leading--twist expression Eq.~(\ref{coef}).

Fitting directly  Eq.~(\ref{tmc}) to the data is impractical, because
 the nonsinglet quark distribution to be determined appears in the
 $I_2$ integral.  Rather, it is more convenient to view the
 contribution proportional to $I_2$ in Eq.~(\ref{tmc}) as a correction
 to be applied to the data. Namely, we re-express $F_2^{\rm LT}$ in the
 $I_2$ integral in terms of the full $F_2$, i.e., to the
 next-to-leading twist level we simply replace $F_2^{\rm LT}$ 
in $I_2$ with $F_2$. 
Therefore, in practice, the function
\be
\label{tmc2}
\Phi_2(x,Q^2) \equiv
\frac{x^2}{\tau^{3/2}}\frac{F_2^{\rm LT}(\xi_{\rm
    TMC},Q^2)}{\xi^2_{\rm TMC}}
\ee
is fitted to the corrected data:
\be
\label{tmc1}
F_2^{\rm \dat-TMC}(x,Q^2)\equiv F_2^{\dat}(x,Q^2)-
6\frac{M^2x^3}{Q^2\tau^2} \int_{\xi}^1 \frac{dz}{z^2}F_2^{\dat}\lp
z,Q^2\rp .
\ee 
The integral over the experimentally measured
structure function $F_2^{\dat}$ can be determined using the
interpolation of the data based on neural networks which was
constructed in Ref.~\cite{f2ns}. Each pseudo-data point is then
corrected using Eq.~(\ref{tmc1}), and the fit procedes as outlined in
the previous sections, with $F_2$ replaced by $\Phi_2$.

Further higher-twist corrections are due to the contribution of 
subleading twist operators in the Wilson expansion. 
A possible
next-to-leading twist contribution from these terms can be
parametrized  as
\be
\label{htparm}
F_2^{\rm NLT}(x,Q^2)=F_2^{\rm LT}(x,Q^2)\left(1+\frac{HT(x)}{Q^2}\right) .
\ee
We will then assess the size of these corrections by parametrizing
the function $HT(x)$ with a neural network and fitting it to the data,
as we will discuss  in Sect.~\ref{ht}. 

\clearpage

\section{Neural networks}
\label{netwprks}

Neural networks, which we use in order to parametrize parton distributions,
are nonlinear maps between input
$\xi_i^{(1)}$ and output $\xi_i^{(L)}$ variables, i.e. they are just a
particularly convenient set of interpolating functions. 
Like other sets of 
functions (such as e.g. orthogonal polynomials), neural networks, in the
limit of infinite 
size, can
reproduce any continuous function. However, the standard 
sets of basis function impose, upon truncation, a
specific bias on the form of the fitted function: for example, polynomials of
fixed degree can only have a maximum number of nodes and stationary
points, and so on. This makes it difficult to obtain an unbiased
representation of an unknown function. 

Whereas clearly finite-size neural networks also impose a limitation
on the set of functions that they can reproduce, their nonlinear
nature makes it possible to  make sure that this is not a source of
bias. In particular, it is possible to show that stability of the fit
is obtained, in that increasing the size of the neural network the
results of the fit do not change. Very crudely speaking, this is a
consequence of the fact that the smoothness of the fitted function
decreases during the fitting process. As a consequence, it is possible
to stop the fitting procedure before one attains 
the lowest value of the $\chi^2$
(or other figure of merit), when the fit is optimized according to
suitable statistical 
stopping  criterion, but before statistical fluctuations are
fitted. It is then possible to check that the result is independent of
the size of the neural network: eventually, as the size increases, the
best-fit result
becomes independent of it. This is to be contrasted with standard
fits, where one reaches the lowest $\chi^2$ compatible with the given
functional form, and eventually if one increases the size of the
fitting function no stable fit can be obtained. 

The critical aspects
of the construction of a neural network parametrization are thus its form,
its fitting (also called training) algorithm, 
and the criterion which is used to stop the training at the best
fit. We now discuss in turn these points, by presenting both the
general strategy and specific choices.

\subsection{Structure}
\label{struct}

We use multi-layer feed-forward neural networks, schematically
depicted in Fig.~\ref{nn}. In these neural networks, 
 neurons are arranged into layers $l=1,\ldots,L$, with
$j=1,\dots,n_l$ neurons per layer. The output is given by neurons in
 the last layer, as a function of the output of all neurons in  the
 previous layer, which in turn is a function of the output of all neurons in  the
 previous layer and so on, starting from the first layer, which provides the
 input. The output $\xi_j^{(l)}$ 
of each neuron ($j$-th neuron of the $l$-th layer)
is given by a nonlinear activation function
 $g(x)$:
\be 
\xi^{(l)}_i=g\lp h_i^{(l)}\rp \ , \quad i=1,\ldots,n_l \ ,\quad\ 
l=2,\ldots,L ,
\label{m1}
\ee
evaluated as a linear combination  of the output $\xi_j^{(l-1)}$ of
all networks in the previous layers, 
\be
\label{m2}
h_i^{(l)}=\sum_{j=1}^{n_{l-1}}\omega_{ij}^{(l)}\xi_{j}^{(l-1)} -\theta_i \ , 
\ee 
where $\omega_{ij}$ (weights) and  $\theta_i$ (thresholds) are free
parameters to be determined by the fitting procedure, and 
$g(x)$ is taken to be a sigmoid in the
inner layers, 
\be 
\label{acfun}
g(x)=\frac{1}{1+\exp(-x)} \ , 
\ee
and linear $g(x)=x$ for the last layer.
\FIGURE[ht]{\epsfig{width=0.8\textwidth,figure=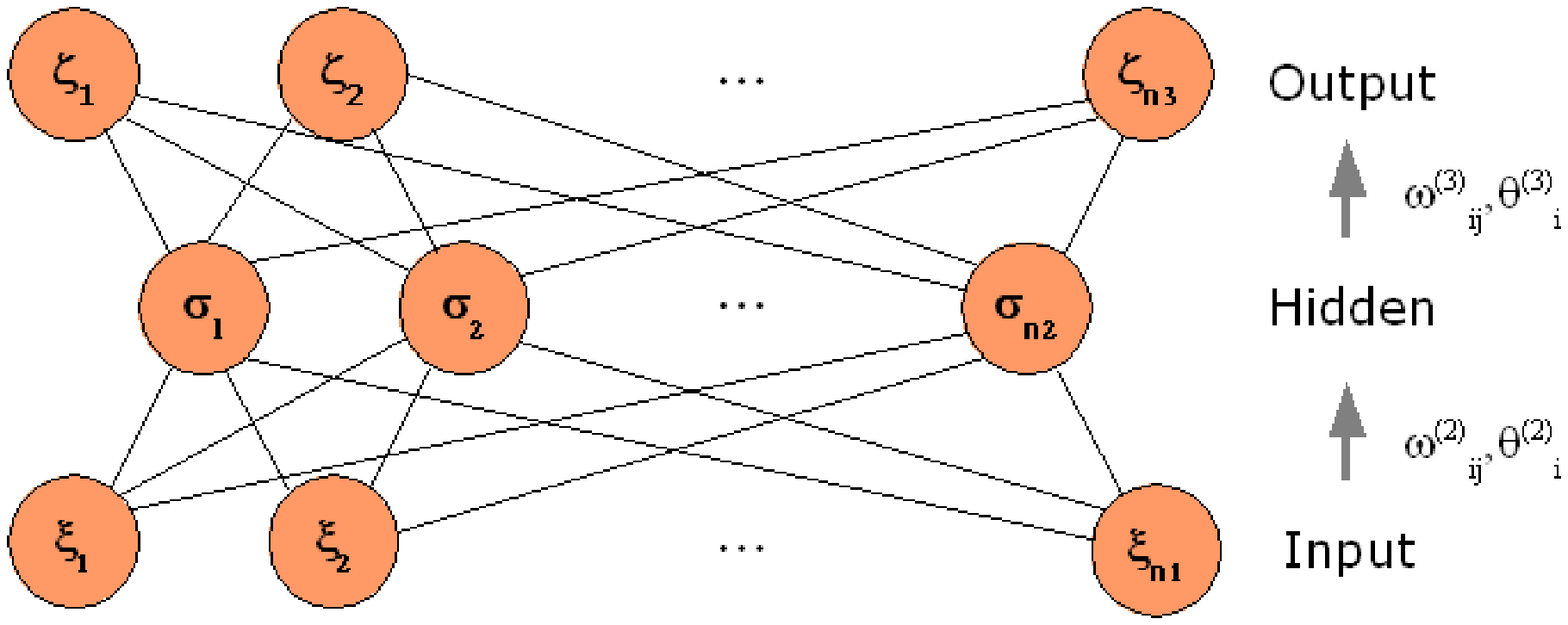} 
  \caption{Schematic diagram of a feed-forward neural network. }
  \label{nn}
}
In practice, it turns out to be convenient to rescale both the input
and the output of the neural network, in such a way that they both take
values between $0$ and $1$: this avoids having weights $\omega_{ij}$
whose numerical values span many orders of magnitude. We have
explicitly checked independence with respect to reasonable variations of
this rescaling.

In short, the neural network outputs the values $ \xi_{j}^{(l)}$ as a
function of the input values $ \xi_{j}^{(1)}$ and the parameters
$\omega_{ij}$, $\theta_i$. The training of the neural network consists of
the determination of the best--fit values of these parameters given a
set of input-output patterns (data).

As discussed previously in detail~\cite{f2ns}, the choice of the architecture
of the neural network cannot be derived from general rules and it must be
tailored to each specific problem. One can roughly guess the size of
the neural network based on rules of thumb, then verify by actual fitting
which is the critical size above which results become  independent
of the size of the networks. Finally, one chooses a size which is
somewhat larger than the critical one, in order to make sure that
results are unbiased. 

In previous work~\cite{f2ns}, it was found that
a neural network with two hidden layers and architecture 4-5-3-1 is
adequate for a fit of the nonsinglet structure function
$F_2(x,Q^2)$. The neural network has four inputs because it turns out
to be more efficient to take simultaneously as input both $x$, $Q^2$,
and their logs $\ln x$, $\ln Q^2$. In the present case, the neural network
only fits the nonsinglet quark distribution at the initial scale, so
only two input neurons for $x$ and $\ln x$ are required. We then
retain the same large architecture, which is surely redundant given
that now only a function of a single variable is fitted. This leads us
to the architecture 2-5-3-1.  Stability of the results upon this
choice will be tested in Sect.~6.3.

The neural network with this structure can be used to parametrize the
nonsinglet quark distribution directly. However, it turns out to be
convenient to actually relate the neural network to the quark distribution
through a suitable factor. This can be viewed as preprocessing of the
data: if the function to be fitted is known to be dominated by some
behavior in a given region, it may be convenient to only fit the
deviation from this behavior.

In our case, it is known that the PDF has to
satisfy the kinematical constraint 
$q_{\NS}(x=1,Q_0^2)=0.$
 Standard counting rules~\cite{rob} as well as existing 
parton fits further suggests that the structure function drops as a
power as $x\to 1$ (up to logarithmic corrections): typically
$q_{\NS}(x=1,Q_0^2)\toone{x} (1-x)^3.$ This suggests that it might be
useful to divide out this expected leading behavior as $x\to1$. The 
vanishing constraint at $x=1$ is
then implemented automatically, which is more efficient than the
solution adopted in previous work~\cite{f2ns,f2p}, where the
constraint was adopted by adding a Lagrange multiplier, i.e., in
practice,
artificial points at $F_2(x=1,Q^2)=0$ with various values of $Q^2$.

Furthermore, the output of the neural network is by
construction bounded. The last layer is a linear function of its
input, which in turn has as its largest value the sum of weights,
given that $\xi_j^{(n_l-1)}$ in turn are bounded,  $\xi_j^{(n_l-1)}\le
1$, according to 
Eq.~(\ref{acfun}). However, both current parton fits as well as
Regge theory  arguments~\cite{rob}
suggest that $q_{\NS}(x,Q_0^2)$ blows up as $x\to 0$. Even though, of
course, this growth is cut off by the fact that the data only extend
down to a minimum value of $x$, and not to $x=0$, this behavior
leads to
substantial variation of the PDF in the measured region as $x$
decreases, and it is therefore advantageous to divide it out.

Therefore, we parametrize the nonsinglet parton distribution as
\be
\label{nntoq}
q_{\NS}(x,Q_0^2)=\frac{(1-x)^3}{x}{\rm NN}(x) \ ,
\ee 
where ${\rm NN}(x)$ is the neural network discussed above. Independence of
the results on the various choices discussed in this section, in
particular architecture and pre-processing, will be checked
explicitly and discussed in the next section.

\subsection{Training}
\label{train}

The procedure of fitting (or training, as it is usually called in the
context of neural networks) is
based on the minimization of a suitable figure of merit. As discussed
in Sect.~\ref{genstrat}, we fit to each replica of the data
a nonsinglet quark distribution by maximum likelihood, i.e. we minimize
the error function
\be
\label{er3}
E^{(k)}\left[{\bf\omega}\right]=\frac{1}{N_{\dat}}\sum_{i,j=1}^{N_{\dat}}
\lp F_i^{(\art)(k)}-F_i^{(\net)(k)}\rp
 \lp \lp 
\mathrm{\overline{cov}}\rp^{-1}\rp_{ij}
\lp  F_j^{(\art)(k)}-F_j^{(\net)(k)} \rp  ,
\ee
%
where $F_i^{(\net)(k)}$ is determined in terms of the nonsinglet quark
distribution at the reference scale $Q_0^2$ by Eq.~(\ref{f2conv}), and
the quark distribution at the reference scale is given by a neural
network Eq.~(\ref{nntoq}). The covariance matrix
$\mathrm{\overline{cov}}^{(k)}$ is defined in Eq.~(\ref{covmatnn}) and
it does not include normalization errors, as discussed in
Sect.~\ref{mcgen}.  Note that $E^{(k)}$ is a property of each
individual replica, whereas the quality of the global fit is given by
the $\chi^2$ computed from the averages over the sample of trained
neural networks, namely
\be
\label{chi2tot}
\chi^2=\frac{1}{N_{\dat}}
\sum_{i,j=1}^{N_{\dat}} \lp F_i^{(\exp)}-\la
F_i^{(\net)}\ra_{\rep}\rp \lp \lp \mathrm{cov}\rp^{-1}\rp_{ij} \lp
F_j^{(\exp)}-\la F_j^{(\net)}\ra_{\rep} \rp ,  \ee 
where now  the covariance matrix, defined in Eq.(~\ref{covmat}),
includes normalization uncertainties.

The error function Eq.~(\ref{er3}) can be minimized with a variety of
techniques, including standard steepest--descent in the space of
parameters. The back-propagation method, which was used in
Refs.~\cite{f2ns,f2p} and is often advantageous for neural network
training, cannot be adopted here because the measured quantity (the
observed value of the structure function) depends non-locally on the
neural network, i.e. it depends on the value of the parton
distribution for several values of the input variable $x$.
It turns out to be convenient to use a genetic
algorithm~\cite{gathesis} as a
minimization method, as already done in Refs.~\cite{f2p,tau}. 

The genetic algorithm applied to our problem works in the following
way, for each replica and repeating the whole procedure for each
replica, 
i.e. 
$N_{\rep}$ times (we shall omit  for simplicity
the index $k$ which identifies the individual
replica). 
The
state of the neural network is represented by the weight vector
\be
\label{chain}
{\bf \omega}=\lp \omega_1,\omega_2,\ldots,\omega_{N_{\parr}}\rp \ .  
\ee 
where each element $\omega_i$ corresponds to a weight
$\omega^{(l)}_{ij}$ or threshold $\theta_i$, and
$N_\parr$ indicates the total number of weights nd thresholds.  
 A set of $N_{\mut}$
copies of the state vector is then generated.
 Minimization is performed in steps usually
referred to as generations (training cycles in ref~.\cite{f2p}). 
At each generation, a new set  of copies of the 
weight vectors is obtained 
by means of  two operations. First (mutation) a randomly chosen 
element of the state vector ${\bf \omega}$) is replaced by a new
value, according to the rule
\be
\omega_k \to \omega_k + \eta \lp r-\frac{1}{2} \rp \ ,
\ee
where $r$ is a uniform random number between 0 and 1 and $\eta$
(mutation rate) is a free parameter of the minimization algorithm
which can be optimized either for the given problem, or dynamically
during the minimization. Second (selection) a set of vectors with low
values of the error Eq.(~\ref{er3}) are selected out of the total
population of $N_{\mut}$ individuals, and use to replace the original
vector. The simplest option is to select the
$N_{\mathrm{sel}}<N_{\mut}$ vectors with lowest error. Methods based
on probabilistic selection, such as those used Ref.~\cite{f2p},
explore more efficiently the space of possible mutations. Moreover, in
order to avoid the local minima and increase the training speed, we
have introduced multiple mutations. Specifically, we have found that
one additional mutation with probability 50\% and two additional
mutations with probability 20\% produce a significant improvement of
the convergence rate.

The procedure is iterated until the vector with smallest value of the
error function of the set for each generation meets a suitable
convergence criterion, to be discussed in the next subsection.  Note
that at each generation the value of the error function never
increases.  The main advantage of genetic minimization is that it
works on a population of solutions, rather than tracing the progress
of one point through parameter space. Thus, many regions of parameter
space are explored simultaneously, thereby lowering the possibility of
getting trapped in local minima.
Usually, this procedure is further supplemented by
crossing within the chains of mutated parameters (see e.g. Ref.~\cite{f2p}).
However, this is not necessary in our case since, thanks
to a dedicated preprocessing, mutations alone are
enough to reach satisfactory minima. Moreover, the
use of Monte Carlo replicas should avoid hitting the same
minimum twice.

Here, genetic minimization is applied  
to an initial set of parameters chosen at
random. In fact, in order to fully explore the space of parameters, it
turns out to be advantageous to first choose at random the range of
parameters, and then their values in this range. In practice,
the range is chosen  at random between 
\be 
\lc-\la \omega \ra-\sigma_{\omega}, \la \omega \ra+\sigma_{\omega}\rc \ , 
\ee
and
\be 
\lc -\la \omega \ra+\sigma_{\omega}, \la \omega \ra-\sigma_{\omega}\rc \ ,
\ee 
where $\la \omega \ra$ and $\sigma_{\omega}$ are the average and
variance of the weights computed from the set of best-fit networks
obtained in a previous fit.  The value of the mutation rate is chosen
to be $\eta=8$, and the probabilistic algorithm of Ref.~\cite{f2ns} is
adopted. While these choices are optimized to obtain fast convergence,
we have checked that they do not influence the final results.

Finally, because we have to deal with data sets coming from different
experiments and with different features, it turns out to be
advantageous to adopt  weighted training, in order to ensure that the
fit to all experiments is of comparable quality.  To this 
purpose~\cite{f2p}, more weight is given to 
experiments with a larger value of $E$. Namely, the  function
$E_{\mathrm{min}}$ to be minimized  is given by
\be
E_{\mathrm{min}}=\frac{1}{N_{\dat}}\sum_{j=1}^{N_{\rm exp}}
p_{j} N_{\dat,j} E_j \ , 
\ee 
where $N_{\dat,j}$ is the number of data points and $E_j$ the value of the
error function defined in Eq.~(\ref{er3}) but restricted to the points coming
from the $j-th$ experiment, and the weights $p_j$ are adjusted
dynamically
according to
\be p_j= \lp
\frac{E_j}{E_{\mathrm{max}}}\rp , 
\label{reweight}
\ee
where $E_\mathrm{max}$ is the highest amongst the values of $E_j$.  In
order to avoid introducing artificial instabilities, the re-weighting
is only used if the ratio on the right--hand side of
Eq.~(\ref{reweight}) differs significantly from one. In practice, no
reweighting is implemented if $c_{\rm min}<E_{NMC}/E_{BCDMS}<c_{\rm
max}$, with $c_{\rm min}=0.78$ and $c_{\rm max}=1.22$.

\subsection{Stopping}
\label{stopp}

As we explained already, the crucial feature which guarantees a bias-free fit
is the possibility of stopping the training not at the lowest value of
the figure of merit (which might depend on the functional form of the
fitting function), but rather  when suitable criteria are met. 
These criteria should single out the point where the fit reproduces
the information contained in the data, but not the statistical noise.
Namely, the best fit should have
the lowest possible value of the figure of merit
compatible with the requirement of not overlearning, i.e. not fitting
statistical fluctuations. In previous work~\cite{f2ns,f2p}, this was
done by determining an optimal training length. The present case is
more subtle, because on the one hand
the fitted
function is much more constrained by the data, on the other hand, the
quantity which is being fitted is not directly observable.
This makes it harder to
distinguish overlearning from an actual improvement in the fit.

We therefore adopt the following criterion. We 
 separate the data set into two
disjoint sets. We then
minimize the error function, Eq.~(\ref{er3}) 
computed only with the data points
of the first set (training set, henceforth). During the minimization
 process, we compute the error function from the data of the second
 set (validation set, henceforth). The best fit has been reached when
the validation error
function ceases to decrease. Overlearning, in particular, corresponds
 to a situation where the training error function keeps decreasing
 while the validation one does not, thereby signaling that the fit is
merely reproducing the fluctuations of the specific training set.
The procedure is reliable and it does not lead to loss of
 information if both the training and the validation set reproduce the
 features of the full data set. This can be simply achieved on average
 by choosing a different random partition for each replica of the
 data, and then checking that the number of replicas is sufficiently large.
Similar methods have been widely used in various applications of neural
networks, and in high--energy physics e.g. in Ref.~\cite{jets}.

In practice, we implement this stopping criterion as follows. 
 We define a training fraction $f_{\mathrm{tr}}=N_{\mathrm{tr}}/
N_{\mathrm{dat}}$ (with default value $f_{\mathrm{tr}}=0.5$) and 
for each replica we select at random a fraction  $f_{\mathrm{tr}}$ of
points for each experiment, which we use for training, while
the remaining points are assigned to a validation set. We then 
  compute separately the
corresponding training and validation covariance matrices. The
training error function $E^{(k)}_{\mathrm{tr}}$  Eq.~(\ref{er3})
is  minimized using a genetic algorithm, and at each generation of
 the genetic minimization the validation error function 
$E^{(k)}_{\mathrm{val}}(l)$ is computed as
 a function of the generation index $l$. Having chosen a fixed
 fraction of points separately for each experiment allows for weighted
 training. 
Both $E^{(k)}_{\mathrm{tr}}$
 and $E^{(k)}_{\mathrm{val}}(l)$ are computed neglecting cross--correlations
 between data in the validation and the training set: we have verified
 that this is a small correction anyway.

We then let the training proceed at least  until $E^{(k)}_{\mathrm{tr}}$ has
  reached the threshold value 
  $E^{(k)}_{\mathrm{tr}}= E_{\mathrm{min}}$. We choose as  
  default the value $E_{\mathrm{min}}=3$ . This ensures that the
  training does not stop due to fluctuations in the early stages.
Beyond this point, i.e. if
$E^{(k)}_{\mathrm{tr}}\le E_{\mathrm{min}}$, the training is
  stopped at the $l$-th generation if the training error function
  decreases,
\be
\frac{\la E_{\rm tr}(l)\ra}{\la E_{\rm tr}(l-N_{\rm sm})\ra} < 1
\label{con1b}
\ee
while the validation error function
\be
\label{con1}
\frac{\la E^{(k)}_{\mathrm{val}}  \lp l \rp\ra}{\la E^{(k)}_{\mathrm{val}} 
 \lp l
-N_{\mathrm{sm}} \rp\ra} \ge 1 \ ,
\ee
where 
\be 
\la E^{(k)}_{\mathrm{val}} \ra \lp l \rp\equiv \frac{1}{N_{\mathrm{sm}}}
\sum_{i=l-N_{\mathrm{sm}}+1}^{l} E^{(k)}_{\mathrm{val}}(i) \ , 
\ee 
with $ E^{(k)}_{\mathrm{val}}(i)$ the validation error function for
the $i-$th GA generation, and similarly for the training error.  In
other words, the training is stopped if the value of the validation
error function averaged over $N_{\mathrm{sm}}$ generation starts
increasing, while the training error function decreases.  This last
condition must be imposed (despite the fact that with a genetic
algorithm the figure of merit always decreases) because of weighted
training: when the weighting is readjusted the error function could
increase locally.  We take $N_{\mathrm{sm}}=4$ as a default value.
This averaging of the figures of merit along the training is 
analogous to the determination of moving averages,
widely used especially in financial dynamics.

Finally, we have introduced an upper length to the training process,
i.e. a maximum number of generation $N_{\mathrm{gen}}$, 
chosen to correspond to a very long training such that the validation
error function is no longer expected to decrease. 
 The default value is
  $N_{\mathrm{gen}}=800$. In practice, the criterion
 condition~(\ref{con1}) was always met well before this point except
 in a tiny number of cases.

\section{Results}
\label{results}

We present our best-fit result for the nonsinglet quark
distribution and its statistical features, and in particular we 
show that it provides a consistent estimate of both central values and
errors. We will then show its stability upon variation of the fitting 
procedure.

\subsection{Next-to-leading order results: central values}
\label{nlocent}

\FIGURE{
\epsfig{width=0.71\textwidth,figure=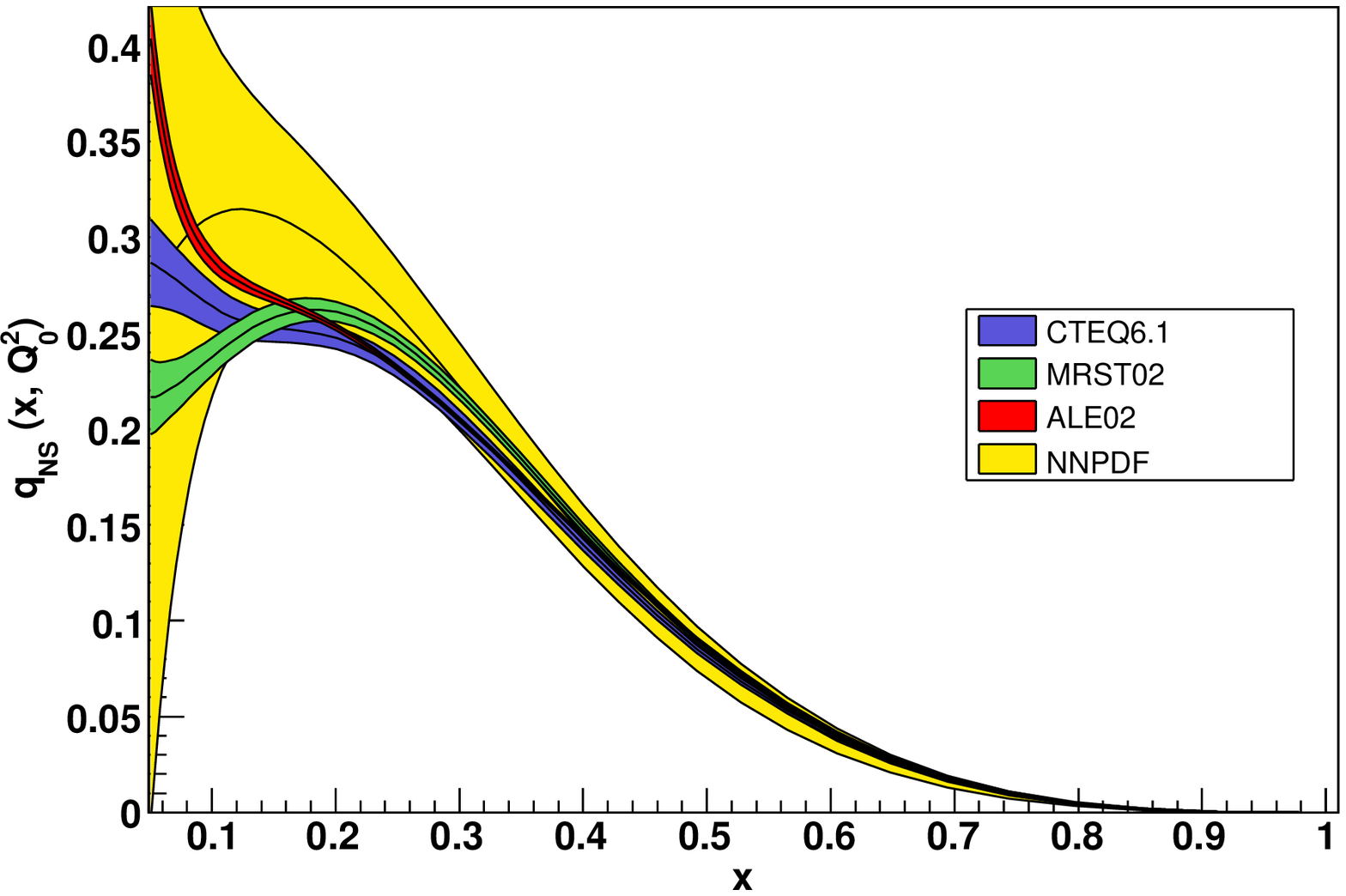} 
\caption{\small The best-fit NLO nonsinglet quark distribution $q_{\rm
    \NS}(x,Q_0^2)$ in the large-$x$ region. The
  MRST, CTEQ and Alekhin determinations are also shown for
  comparison. In this and subsequent plots of $q_{\rm
    \NS}$ we take $Q^2=Q_0^2=2$~GeV$^2$.}
\label{finq1}
}

\FIGURE{
\epsfig{width=0.71\textwidth,figure=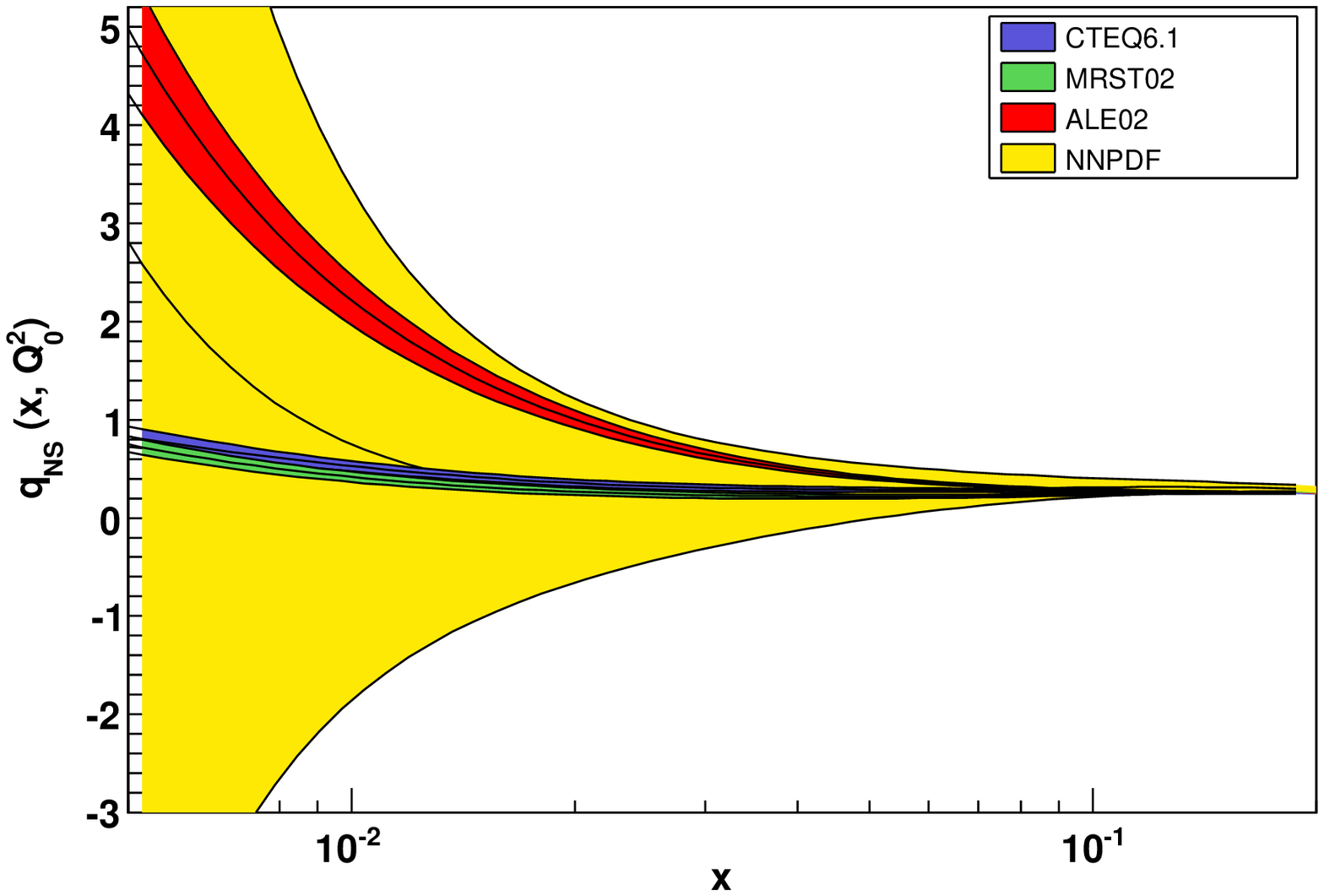} 
\caption{\small The best-fit NLO nonsinglet quark distribution $q_{\rm
    \NS}(x,Q_0^2)$ in the small-$x$ region. The
  MRST, CTEQ and Alekhin determinations are also shown for comparison.}
\label{finq2}
}

The result obtained from a set of $N_{\mathrm{rep}}=1000$ neural
networks trained to replicas of the data with the method described
above, using the NLO expressions of the coefficient function and
anomalous dimension is displayed in Figs.~\ref{finq1} and
~\ref{finq2}.  
The error band corresponds to a one--sigma contour.  The
distributions of values of the error function $E$ Eq.~(\ref{er3}) for
the training and validation samples are displayed in
Fig.~\ref{stopchi2}. These distributions are poissonian (gaussian) to
good approximation, and both are peaked around a similar value,
thereby showing that the quality of the fit for points included or not
included in the fit is equally good. The distribution of training
lengths is shown in Fig.~\ref{stoplength}. It is poissonian to good
approximation, and it shows that convergence is reached after a small
number of generations of the genetic algorithm, much smaller than the
maximum $N_\gen$ which is only reached in a tiny number of cases.
\FIGURE{ \epsfig{width=0.45\textwidth,figure=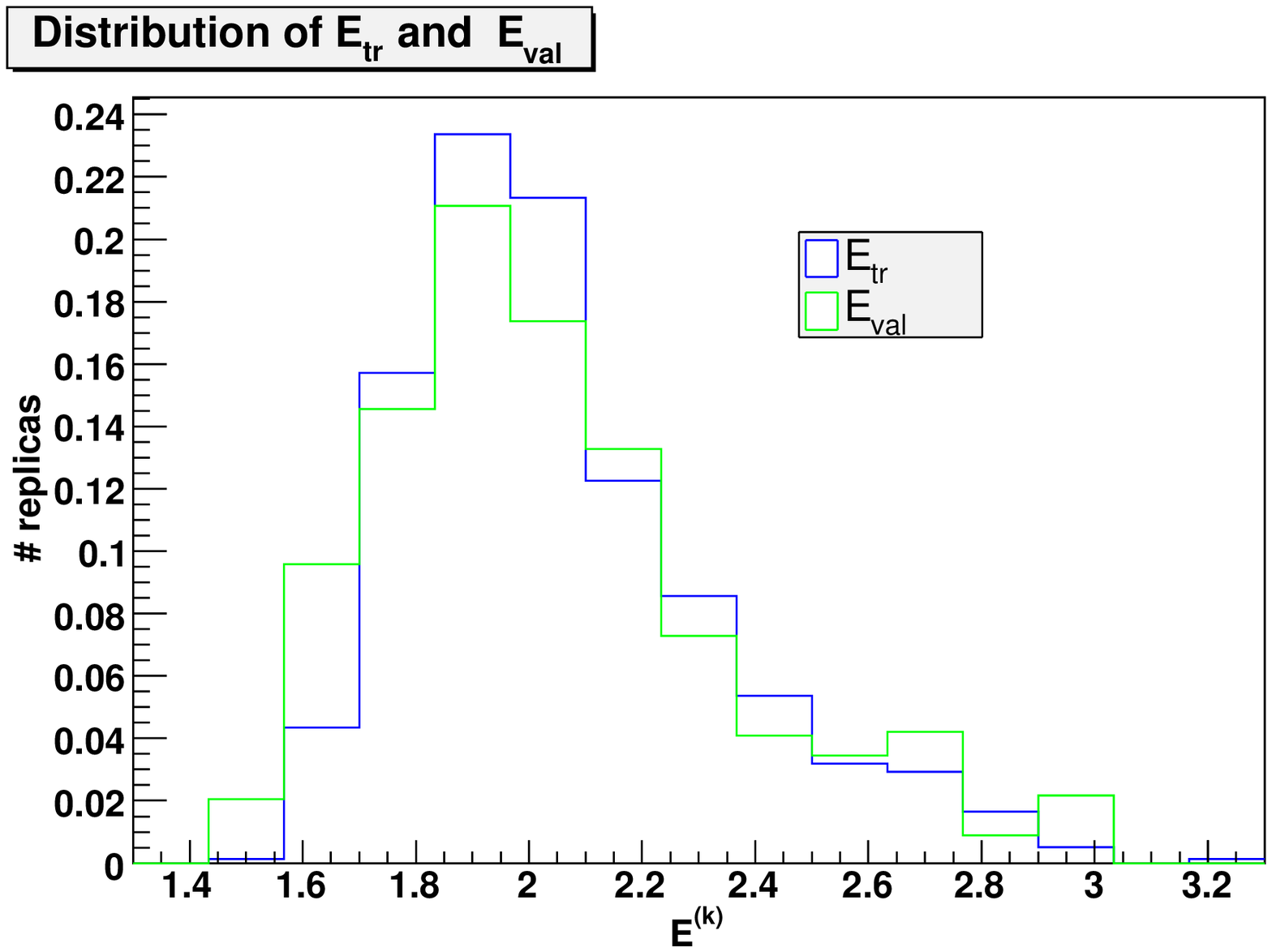}
\epsfig{width=0.45\textwidth,figure=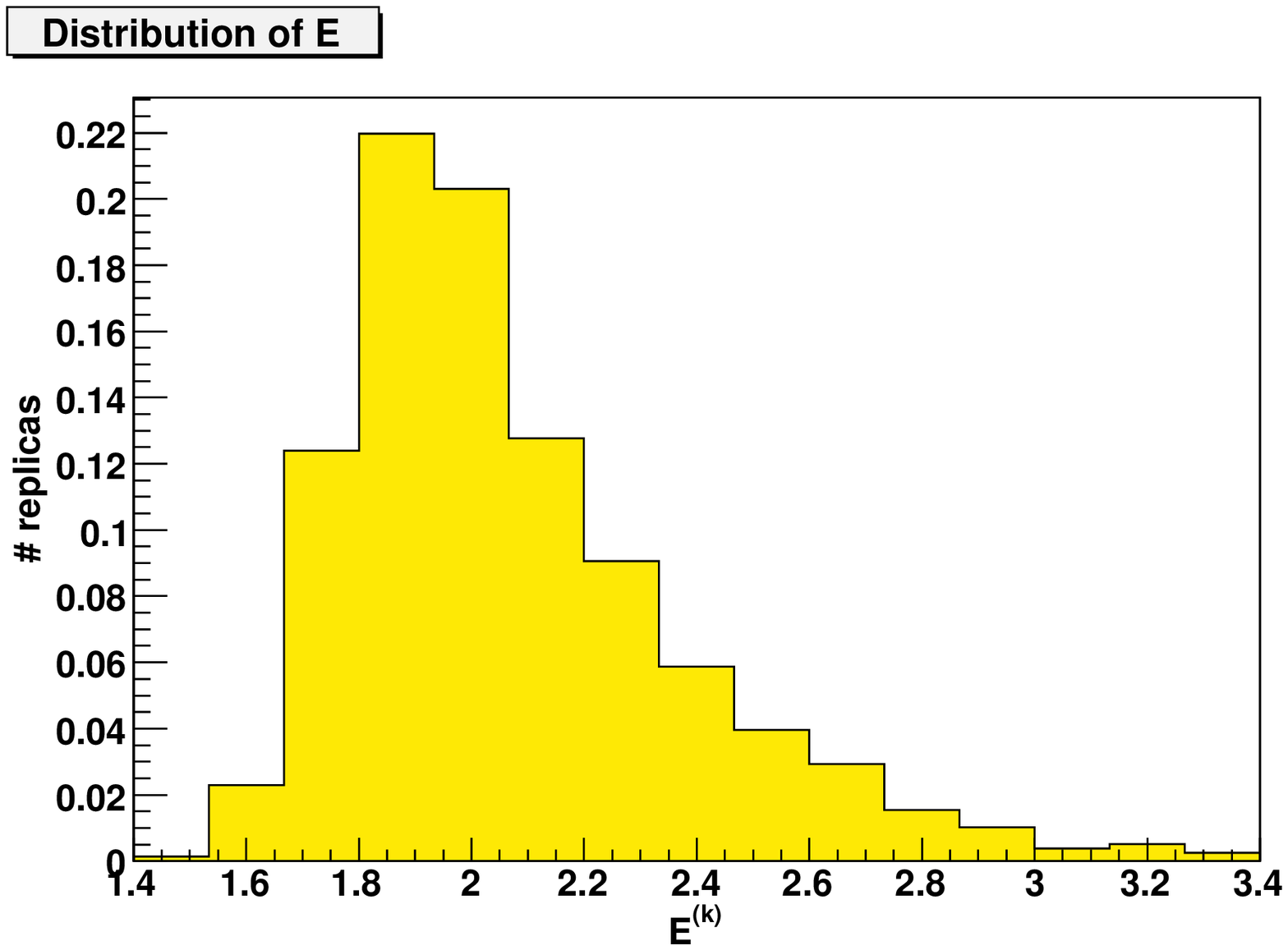}
\caption{\small Distribution of values of the error function $E$
Eq.~(\ref{er3}) at the stopping point for the training and validation
samples (left), and for the full data sample (right).}
\label{stopchi2}
\label{stopchi2t}}
\FIGURE{
\epsfig{width=0.45\textwidth,figure=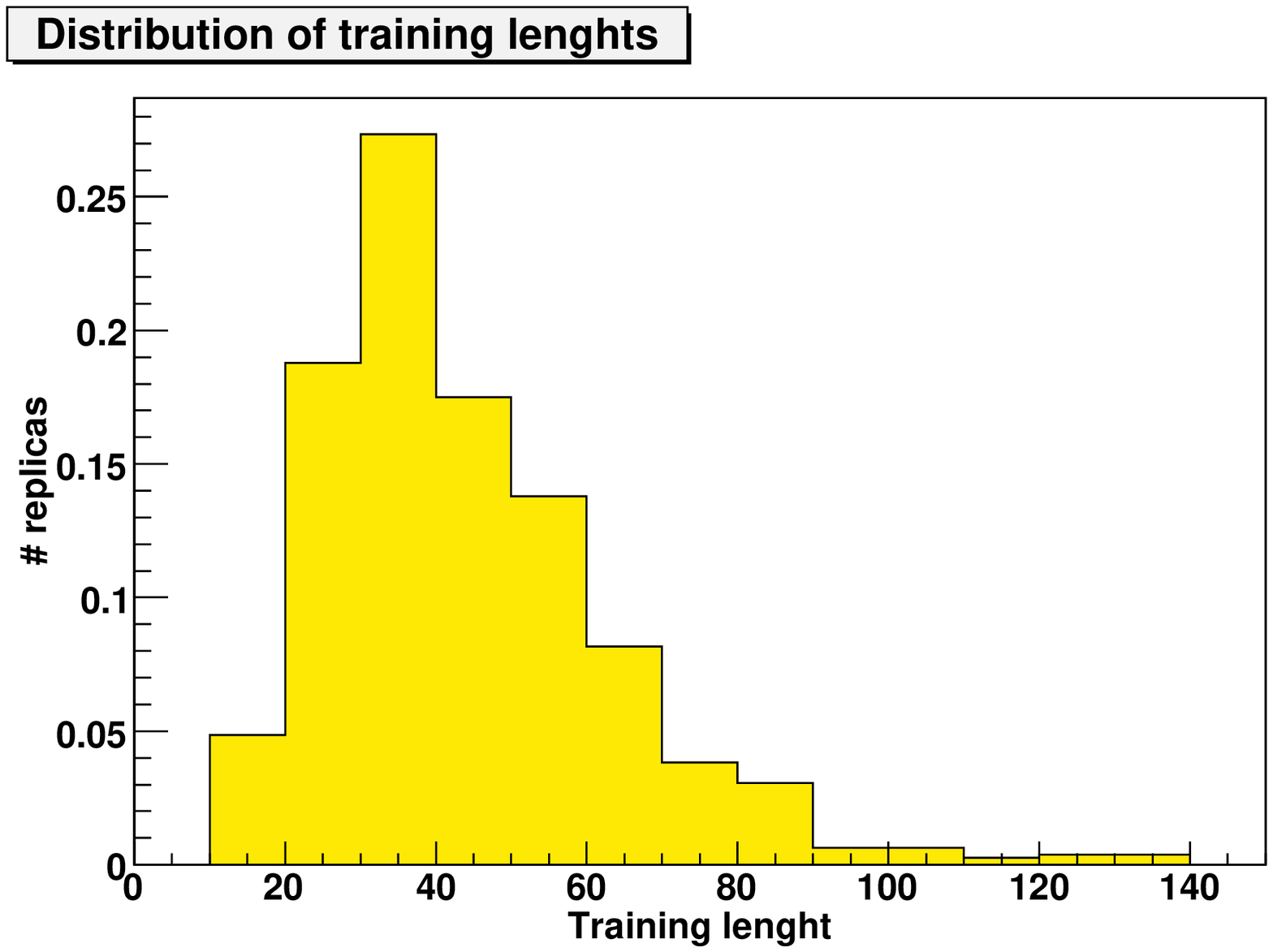}
\caption{\small Distribution of training lengths.}
\label{stoplength}
}

The  distribution of values of the error
function $E$ Eq.~(\ref{er3}) for the  total data set is displayed in
Fig.~\ref{stopchi2t}, while  the
statistical properties of the fit
are summarized in
Table~\ref{est}. Note that, because  the quantities listed in the
table characterize the comparison of the fit to the data 
(see Appendix~\ref{dataest} for the definitions), they are determined from
the structure function $F_2^{\rm \NS}$ rather than the quark
distribution itself.
The error functions of each fit are distributed to
good approximation in a gaussian way about the value $\langle E\rangle=2.27$ given in
Table~\ref{est}. A value   $\langle E\rangle\approx 2$
is expected for a good fit because if errors are correctly estimated, the
average standard deviation of true values about the measured value
should equal the total uncertainty, but  replicas are generated
about the measured values so their average standard deviation about
the true values should be twice the total uncertainty. However, the
fit to each individual replica should be closer to the true value,
because of the availability of many data points at various values of
$Q^2$ which are combined by the fit. Indeed, the average error of the
fit is much smaller than the average experimental error:
$\la \sigma^{(\net)}
\ra_{\dat}\ll \la \sigma^{(\exp)}
\ra_{\dat}$ (see Table~\ref{est}). 

The fact that the figure of merit for the global fit Eq.~(\ref{chi2tot}) 
takes the value $\chi^2\approx 1$  shows that the central value of the
fit, after averaging over replicas,  is distributed about the measured
values as one expects of the correct distribution of true values,
namely, with variance equal to the experimental error (which is much
larger than the error on the fit itself as we just discussed). Note
that,
strictly speaking, the total number of independent degrees of
freedom is somewhat smaller  than the total number of points. However,
the number of effective free parameters of the neural network is rather
smaller than the total number of weights and thresholds,  because our
choice of stopping criterion ensures that the neural network is
redundant. Hence, the number of degrees of freedom is at most by a few
percent smaller than $N_\dat$ (and the correctly estimated $\chi^2$
accordingly larger), which is smaller than the uncertainty on $\chi^2$
due to statistical fluctuations, which is
of order of $N_{\dat}^{-1/2}$. However, as a consequence of this, we
expect the value of $\chi^2/N_\dat$ to be systematically lower than
one by a few percent for an optimal fit. This expectation will be
borne out by our results.

\TABLE[ht]{
\begin{tabular}{|c|c|c|c|}
\hline
& $\quad$ Total $\quad$ & $\quad$ NMC
$\quad$  & $\quad$ BCDMS $\quad$  \\
\hline 
$\chi^{2}_{\tot}$ & 0.75  & 0.72 & 0.78 \\
$\la E \ra $   &  2.27 & 1.99 & 2.52    \\
\hline
$r\lc F_2^{\NS}\rc$ & 0.81 & 0.66 & 0.95\\
\hline
 $\la \sigma^{(\exp)}
\ra_{\dat}$ & 0.011  & 0.017& 0.006\\
 $\la \sigma^{(\net)}
\ra_{\dat}$&  0.006 & 0.009& 0.004\\
 $r\lc \sigma^{(\net)}
 \rc$ &  0.59 & -0.04& 0.86\\
\hline
 $\la \rho^{(\exp)}
\ra_{\dat}$ & 0.11& 0.39& 0.16  \\
$\la \rho^{(\net)}
\ra_{\dat}$ & 0.46   & 0.42& 0.50\\
$r\lc \rho^{(\net)}\rc$ & 0.15  & 0.25 & 0.04 \\
\hline
 $\la {\rm cov}^{(\exp)}
\ra_{\dat}$& $8.6~10^{-6}$  & $1.0~10^{-5}$ & $7.2~10^{-6}$\\
$\la {\rm cov}^{(\net)}
\ra_{\dat}$&  $2.1~10^{-5}$ & $3.8~10^{-5}$& $6.9~10^{-6}$\\
$r\lc {\rm cov}^{(\net)}\rc$ & 0.24& 0.23& 0.57\\
\hline
\end{tabular}
\caption{\small Statistical estimators for the final
sample  of $N_{\rep}=1000$ neural networks, both for the
total data points and for the individual experiments incorporated
in the fit (NMC and BCDMS).
}
\label{est}
}

We conclude that the central fit is correctly
estimated.

\subsection{Next-to-leading order results: uncertainties}
\label{nlounc}

Because experimental errors are much larger than the error
on the fit, the values of $\la E \ra$ and $\chi^2$ discussed in
Sect.~\ref{nlocent} 
do not
test for the accuracy of the one--sigma error band on the central fit.
 We can  check for the correctness of the latter by  studying  the
fluctuation of predictions obtained from  different sets of replicas.
To this purpose, we note that we can view the prediction $q_i$ obtained for
the central values of the quark distribution by averaging over
$N_\rep$ replicas, as a random variable, whose variance is given by
\be
\label{meanvar}
V[q_i]=\frac{\sigma^2_{q_i}}{N_{\rep}},
\ee
in terms of the variance $\sigma^2_{q_i}$ of the set of
replicas. We can define an average distance $d[q]$ as the
root--mean square difference between values of $q_i$ obtained by
averaging over different set of replicas, normalized to the variance
Eq.~(\ref{meanvar}) (see appendix B). 
If the error on $q$ is correctly
estimated, the average of $d[q]$, determined for different points, 
must approach one. 
We can 
similarly view the standard deviation  of each data point
$\sigma^2_{q_i}$, and  test for their statistical accuracy by computing
the corresponding distances. 

\TABLE[ht]{
\begin{tabular}{|c|c|c|c|}
\hline 
$N_{\rep}$ &  10   &  100 
&   500  \\
\hline
$\la d\lc q \rc\ra_{\dat}$  & 1.02& 0.96&   0.92\\
$\la d\lc q \rc\ra_{\extra}$  & 1.11& 0.99&  0.85\\
\hline
$\la d\lc \sigma_q \rc\ra_{\dat}$   & 0.93& 0.88& 0.93 \\
$\la d\lc \sigma_q \rc\ra_{\extra}$   & 1.13& 0.97 & 0.91\\
\hline
\end{tabular}
\caption{\small  Stability estimators as a function of the
number of trained replicas $N_{\rep}$. }
\label{stabtablenrep}
}
Hence, we test for accuracy of
estimates of errors and correlations by computing this distance for
the quark distribution $q_i$ (determined as an average over neural
networks) and  the error on it $\sigma^2_{q_i}$ (determined as variance of the
networks). The distance is determined between results obtained from a
set of $N_{\rm rep}$ replicas each. The
computation  is performed  for 14 points linearly
spaced in $x$ in the data region ($0.05\le x\le 0.75$) 
and
then averaged, and similarly
for the same number of
points logarithmically spaced in the extrapolation region ($10^{-3}\le x\le
10^{-2}$). In order for the result to be significant we must make sure
that its standard deviation $\sigma_d$  is not too
large. However, because $d$ is a standard deviation itself (normalized
to its expected value), $\sigma_d$ only
scales as $n^{-1/4}$ with the number of points $n$, so  $\sigma_d\sim
0.5$. On the other hand,
increasing the number of points will not help because, of course, two
values of $q(x)$ and $q(x^\prime)$ are highly correlated when $x$ and
$x^\prime$ are close to each other. Thus, we obtain stability by
repeating the computation
 for different random choices of two sets of $N_\rep$ replicas among 
the starting $2N_\rep$ ones, and averaging the results, which reduces
the standard deviation according to Eq.~(\ref{meanvar}). We average over 1000
sets, so that $\sigma_d\sim 0.01$. 

Results with increasing values of $N_\rep$ are shown in
Table~\ref{stabtablenrep}. The distance already converges to the
expected value of one, to percent accuracy, when $N_{\rm rep}=10$. Note
that, because of Eq.~\ref{meanvar}, the expected distance decreases
as $\frac{1}{\sqrt{N_{\rm rep}}}$: the stability of the result we get
when the number of
replicas is further increased from 100 to 500 shows that this
behaviour is indeed observed in our sample of neural networks.

\subsection{Stability upon variation of the neural network structure}
\label{stabarch}
After verifying that our results are statistically consistent, we wish
to check that they do not depend on the details of the fitting
procedure. We do this by determining the distance discussed in
Sect.~\ref{nlounc}, but now computed for a pair of fits which differ
in fitting procedure.   
Independence is verified if the distance equals
its value predicted on statistical grounds (i.e. that which would be
obtained from pairs of replicas coming from the same fit).
Specifically, the computation is done by taking
$N_\rep=50$, and then averaging over 1000 different choices of these
50 replicas among a total set of $N_\rep=100$ replicas. Note that,
because of Eq.~\ref{meanvar}, this verifies independence of the
fitting procedure to an accuracy which is by a factor
$\sqrt{50}$ smaller than the uncertainty on the central value.

\TABLE{
\begin{tabular}{|c|c|}
\hline
Architecture & 2-4-3-1  \\
\hline
$\chi^2$ & 0.75\\
\hline
$\la d\lc q \rc\ra_{\dat}$  & 0.9\\
$\la d\lc q \rc\ra_{\extra}$  & 0.9\\
\hline
$\la d\lc \sigma_q \rc\ra_{\dat}$   & 0.9\\
$\la d\lc \sigma_q \rc\ra_{\extra}$   & 1.4\\
\hline
\end{tabular}
\caption{\small Independence of the architecture of
  the neural networks.}
\label{archtable}
}
The first check is to verify that our results are indeed independent of the
neural network architecture. 
We have compared the results of the fit with the reference architecture
2-5-3-1 with the results of a similar fit with a smaller neural network
architecture 2-4-3-1. The result of this comparison can be 
seen in Table~\ref{archtable}.  
 This 
confirms that we have achieved independence of 
the number of parameters used, both in the data and in the
extrapolation region, a property that is very difficult to achieve in
standard parton fits with fixed functional form.

Next, we estimate the dependence of the results on the different kind of
preprocessing, namely, on the values of the exponents $m$ and $n$ 
\be
\label{pdfexp}
q_{\NS}(x,Q^2_0)=\frac{(1-x)^m}{x^n}NN(x)  \ ,
\ee
when varied about
the default values $m=3$ and $n=1$. In Table~\ref{stabtable} we
display the dependence of the $\chi^2$, computed from a set of
$N_\rep=100$ replicas,  and the stability of the fit
when $1\le m\le 4$ (with $n=1$ fixed) and when  $0.5\le n\le 1.25$ (with
$m=3$ fixed). It appears that the fit is reasonably stable if $2.5\le
m\le 3.5$ and  $0.5\le
n\le 1.25$: central
values in the data region fluctuate less than three $\sigma$ while the
uncertainty is stable. The central fit starts changing in a
statistically significant way when $m$ and $n$ are varied outside
these bounds. However, in such case the quality of the fit, as measured
by the $\chi^2$,
deteriorates considerably. Hence, we conclude that the fit is
independent of the choice of $m,n$ provided these are kept within a
range which allows for an optimal quality of the fit.

Fits within the stability region of the preprocessing exponents are
compared in 
Fig.~\ref{ndepsmallx}. Stability of central values
and errors in the data and extrapolation region are apparent. Note
that the distance Eq.~(\ref{meanvar}) is computed with $N_\rep=50$,
hence the distance of the central values equals  about 15\% of the
error band displayed in the plot. 
\FIGURE[ht]{
\epsfig{width=0.47\textwidth,figure=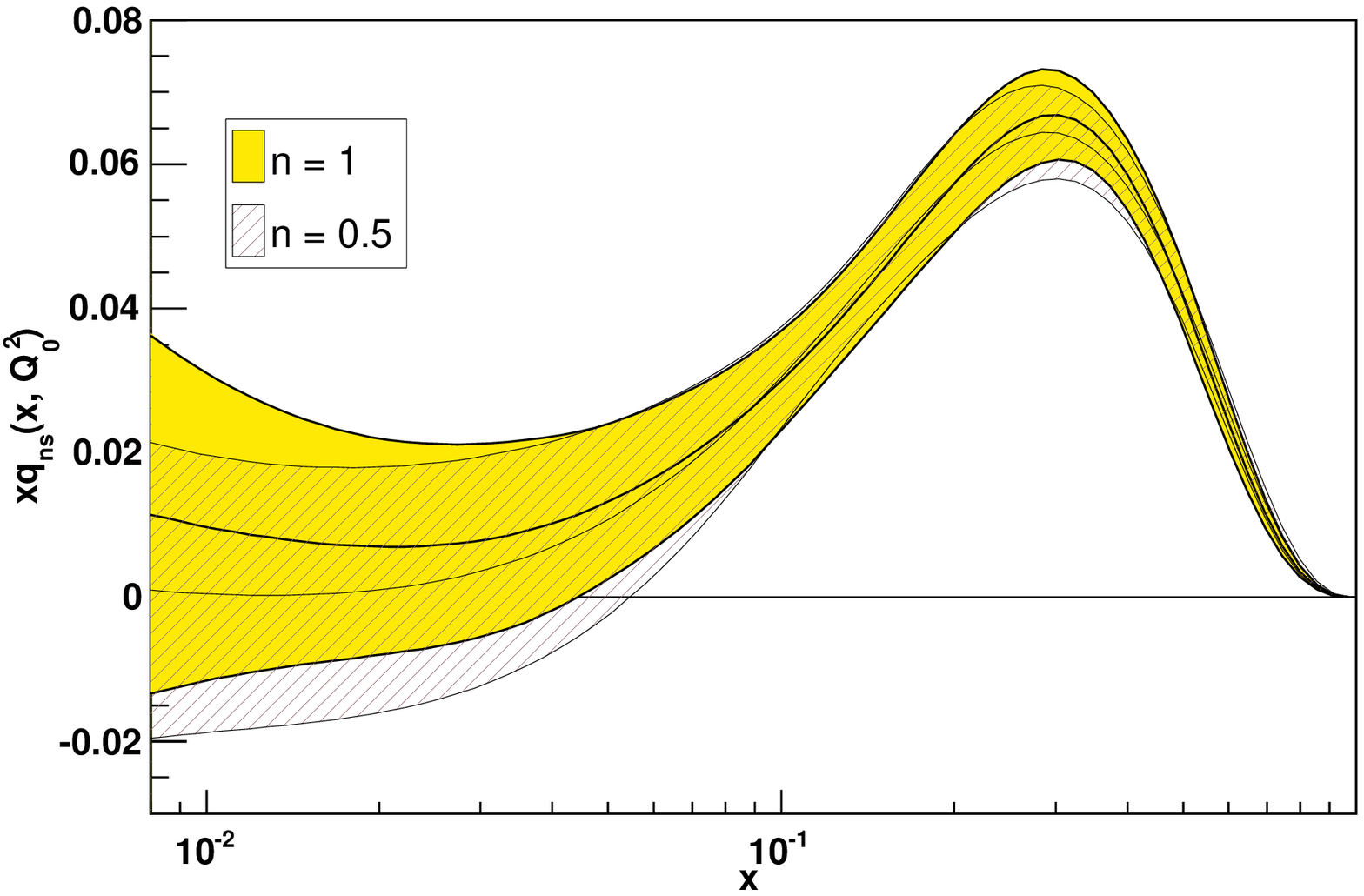}
\epsfig{width=0.47\textwidth,figure=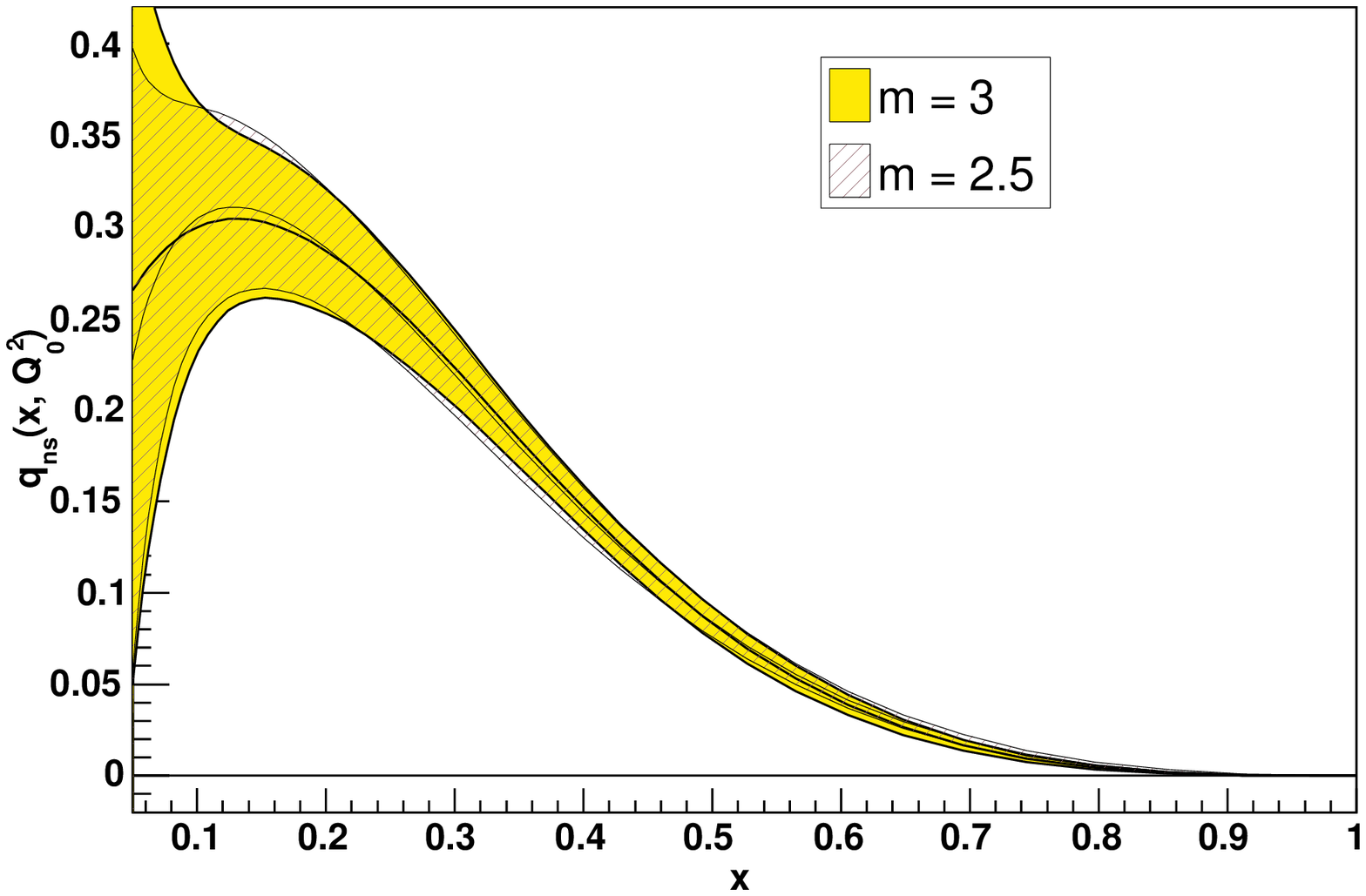}
\caption{Dependence of the results on the preprocessing exponents
$m$, $n$, Eq.~(\ref{pdfexp})\label{ndepsmallx}}
}

\TABLE{
\begin{tabular}{|c|c|c|c|c|c|c|}
\hline
$(m,n)$& (1,1)  & (2,1) & (2.5,1) & (3,1) & (3.5,1) & (4,1) \\
\hline
$\chi^2$ & 0.85 & 0.94 & 0.83 & 0.75 & 0.77 & 0.84\\
\hline
$\la d\lc q \rc\ra_{\dat}$  & 4.5 & 4.6 & 2.9 & -  &3.3& 6.1\\
$\la d\lc q \rc\ra_{\extra}$  & 0.8 & 0.8 & 0.8 & -  & 0.8& 0.8\\
\hline
$\la d\lc \sigma_q \rc\ra_{\dat}$   & 2.0& 1.7& 1.2 & - & 1.2 & 1.5\\
$\la d\lc \sigma_q \rc\ra_{\extra}$   & 1.9& 2.0& 1.3 & - & 1.0& 1.1\\
\hline
\end{tabular}
\begin{tabular}{|c|c|c|c|c|c|c|}
\hline
$(m,n)$ & (3,0.5) & (3,0.75) & (3,0.9) & (3,1) & (3,1.1) & (3,1.25) \\
\hline
$\chi^2$ & 0.79 & 0.77& 0.74 & 0.75 & 0.76 & 0.77\\
\hline
$\la d\lc q \rc\ra_{\dat}$  & 2.5 & 2.1& 1.0 & -& 1.2& 1.6\\
$\la d\lc q \rc\ra_{\extra}$   & 1.2& 0.8& 0.9 &-& 1.9& 1.5\\
\hline
$\la d\lc \sigma_q \rc\ra_{\dat}$   & 1.3& 1.3& 0.9 & - & 1.0& 1.3\\
$\la d\lc \sigma_q \rc\ra_{\extra}$  & 1.0& 1.3& 1.2 &- & 1.8& 2.5\\
\hline
\end{tabular}
\caption{\small  Dependence on
 the preprocessing exponents. }
\label{stabtable}
}

\section{Theoretical uncertainties} 
So far we have checked that our fit reproduces correctly the
information contained in the data, and thus that the error band
computed from it is statistically consistent. A priori, there are further
sources of theoretical error related to use of perturbative QCD in the
fitting procedure: specifically those related to higher order
perturbative corrections and to higher twist terms. 
In this section we will show that these theoretical
errors are in fact negligible.
All tests  in this section, as in Sect.~\ref{stabarch} are done by
computing  the $\chi^2$ from a set of 100 replicas, and the distance by taking
$N_\rep=50$, and then averaging over 1000 different choices of these
50 replicas among a total set of $N_\rep=100$ replicas.

\label{hoht}
\subsection{Dependence on kinematic cuts}
\label{kc}
We first discuss the dependence of results for the nonsinglet parton
distribution $q_{\NS}(x,Q^2)$ on the kinematical cut in $Q^2$. In
Fig.~\ref{kincuts} and Table~\ref{kincuttab} we compare results
obtained when  the value $Q_{\min}^2$ 
is raised to
$Q_{\minn}^2=9$ GeV$^2$, to results obtained with the default 
value $Q_{\minn}^2=3$ GeV$^2$.
Removing the data points in the region 3 GeV$^2 \le Q^2 \le 9$ GeV$^2$
eliminates the bulk of the NMC measurements, i.e. (see
Fig.~\ref{kincov}) essentially all data with $x<0.1$.

As a consequence,
uncertainties increase sizably in the small-$x$ region.
However, results are remarkably stable in the large $x$ region.
In Tab.~\ref{kincuttab} 
we compare results in three different kinematical
regions: where the two fits with different kinematical cuts have data (region
I, $0.2 \le x \le 0.75$), where only the fit with $Q^2_{\minn}=3$ GeV$^2$ has
data (region II, $0.05 \le x \le 0.1$) and finally the region where both fits
extrapolate (region III, $x\le 10^{-2}$).  Not only in region I the
two fits agree completely, but even in region II, whereas the
uncertainty bands increases  considerably, the central fit is
essentially unaffected. This proves the remarkable stability of our results.

\TABLE{
\begin{tabular}{|c|c|c|c|}
\hline 
  &  $\quad$ $x\le 10^{-2}$ $\quad$  & $\quad$ $5~10^{-2}\le x\le 0.1$ 
$\quad$& $\quad$  $0.2\le x\le 0.75$ $\quad$ \\
\hline
$\la d\lc q \rc\ra$  & 0.6 & 0.8 & 0.9  \\
$\la d\lc \sigma \rc\ra$  & 2.7& 3.2 & 1.4  \\
\hline
\end{tabular}
\caption{\small Comparison of fits with different kinematical
cuts. The fit with $Q_{\minn}^2=3~\mathrm{GeV}^2$ (reference) is compared
to the fit with $Q_{\minn}^2=9~\mathrm{GeV}^2$ in three different
regions of $x$.}
\label{kincuttab}
}


\FIGURE[ht]{\epsfig{width=0.70\textwidth,
figure=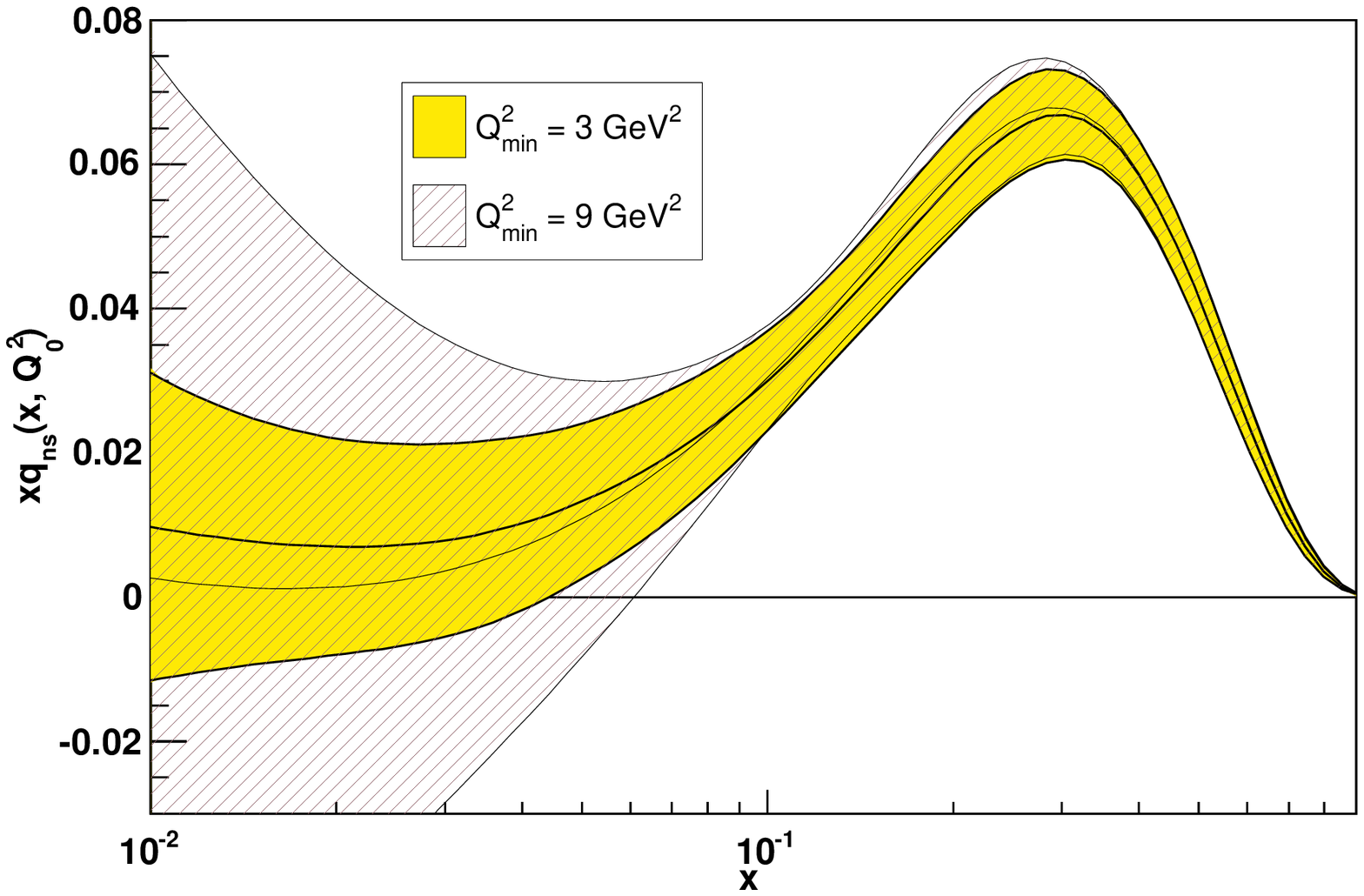} 
        \caption{Results
for $xq_{\NS}(x,Q_0^2)$ with two different kinematical
cuts in $Q^2$.}
        \label{kincuts}}

\subsection{Higher twists}
\label{ht}

\TABLE{
\begin{tabular}{|c|c|c|c|}
\hline
Fit  & $Q_{\min}^2=3~\mathrm{GeV}^2$+HT 
& $Q_{\min}^2=5~\mathrm{GeV}^2$ & $Q_{\min}^2=5~\mathrm{GeV}^2$+HT \\
\hline
$\chi^2$ & 0.76 & 0.79 & 0.78 \\
\hline
$\la d\lc q \rc\ra_{\dat}$  & 2.9 & 0.8 &  3.2 \\
$\la d\lc q \rc\ra_{\extra}$  & 1.4 & 0.8 & 0.9 \\
\hline
$\la d\lc \sigma_q \rc\ra_{\dat}$   & 1.2 & 1.5 & 1.9 \\
$\la d\lc \sigma_q \rc\ra_{\extra}$   & 1.3 & 1.8 & 2.3 \\
\hline
\end{tabular}
\caption{\small  Results of the higher-twist analysis. Stability is
  computed relative to the default NLO fit.}
\label{httab}
}

We now test for the possible role of higher twist terms, on top of the
target--mass corrections which are already included in our default fit
as discussed in Sect.~\ref{tmcht}. To this purpose, we switch on a
twist--four contribution according to Eq.~(\ref{htparm}), and
parametrize $HT(x)$ with a 1-2-1 neural network. In Table~\ref{httab}
we compare the $\chi^2$ obtained by including this higher twist term
to that obtained by raising the cut in $Q^2$. On the one hand, we see
that including higher-twist corrections without raising the $Q^2$ does
not lead to any improvement in the quality of the fit. On the other
hand, we see that if the $Q^2$ is raised, the $\chi^2$ also does not
improve --- in fact it deteriorates slightly, in a way which is
unaffected by the inclusion of higher twist terms.

We take the lack of improvement in $\chi^2$ when either the cut is
raised, or higher twist terms are included, as an indication of the
fact that there is no evidence whatsoever for higher twist terms even
with our lower default cut. The slight deterioration in $\chi^2$ when
the $Q^2$ cut is raised is likely to be related to the fact that experimental
uncertainties in the small--$Q^2$ region (i.e. experimental
uncertainties on NMC data) are somewhat overestimated. 

We conclude that there is no evidence for higher twists for nonsinglet
$F_2^{\NS}$ data with $Q^2>3$~GeV$^2$.

\subsection{Higher orders and   the value of the strong coupling}
\label{hoasval}

So far we have discussed the determination of the nonsinglet parton
distribution through a NLO analysis. We now compare these results to
those obtained at one less and one more perturbative order. Hence, we
produce a full LO fit, with $\amz=0.130$ and a full NNLO fit, with
$\amz=0.115$. The results of these fits are compared to the NLO ones in
Table~\ref{ordertab}.  No variation
of the quality of the fit is found between different perturbative
orders: the values of $\chi^2$ and $\la \sigma^{(\net)} \ra_{\dat}$ are
unaffected by the perturbative
order at which the computation is performed. 
Indeed, only very small improvements of the $\chi^2$ when going
from LO to NLO and NNLO have been found in recent nonsinglet
fits~\cite{grs,bbg}. 

 This means that the effects of higher--order corrections are
entirely reabsorbed in a change of the initial quark
distribution. This change is statistically significant when
going from LO to NLO: the central values moves by $d\approx10$ (see
Tab.~\ref{ordertab}), i.e. by about half the error on the central
value. Indeed,  we have verified that if Tab.~\ref{ordertab} is
recomputed using  a set of $N_\rep=50$ replicas, we get
$\la d\lc q \rc\ra_{\dat}=3.2$ for the LO fit. This means that the
LO and NLO central values differ by a fixed amount, independent of the
number of replicas used to determine the central value, therefore if
we normalize to the standard deviation Eq.~\ref{meanvar} the result
grows as $\sqrt{N_\rep}$.
 No statistically significant difference in central values is observed
 when going from NLO to NNLO. Hence, 
our result supports the conclusion~\cite{grs} that
nonsinglet data are insufficient to provide evidence for higher--order
perturbative corrections. 
 These results are
apparent in Fig.~\ref{order}, where the LO, NLO and NNLO fits are
compared.

\TABLE{
\begin{tabular}{|c|c|c|c|}
\hline
Perturbative order  & LO & NLO & NNLO  \\
\hline
$\chi^2$ & 0.751 & 0.750 & 0.754   \\
\hline
$\la d\lc q \rc\ra_{\dat}$  & 10.2 & - & 2.6 \\
$\la d\lc q \rc\ra_{\extra}$  & 1.2 &-& 1.2  \\
\hline
$\la d\lc \sigma_q \rc\ra_{\dat}$   & 2.2 &-& 3.0   \\
$\la d\lc \sigma_q \rc\ra_{\extra}$   & 1.3&-& 1.3  \\
\hline
 $\la \sigma^{(\net)}
\ra_{\dat}$ & $0.51~10^{-2}$& $0.62~10^{-2}$& $0.49~10^{-2}$ \\
\hline
\end{tabular}
\caption{\small  Quantitative comparison of fits at
different perturbative orders. The distance is computed relative to
the NLO fit using a set of $N_\rep=500$ replicas.}
\label{ordertab}
}

\FIGURE[ht]{\epsfig{width=0.47\textwidth,figure=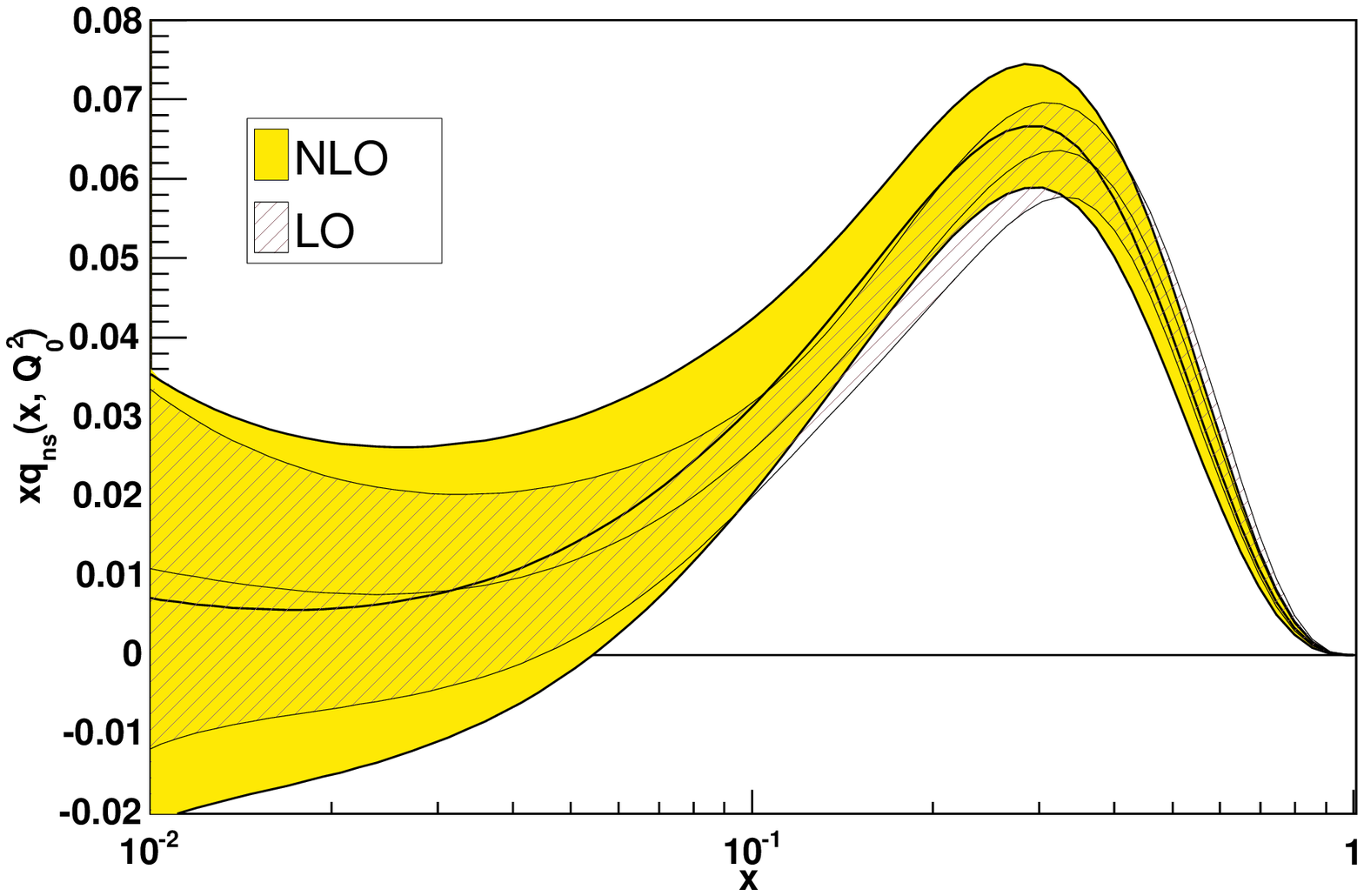}
\epsfig{width=0.47\textwidth,figure=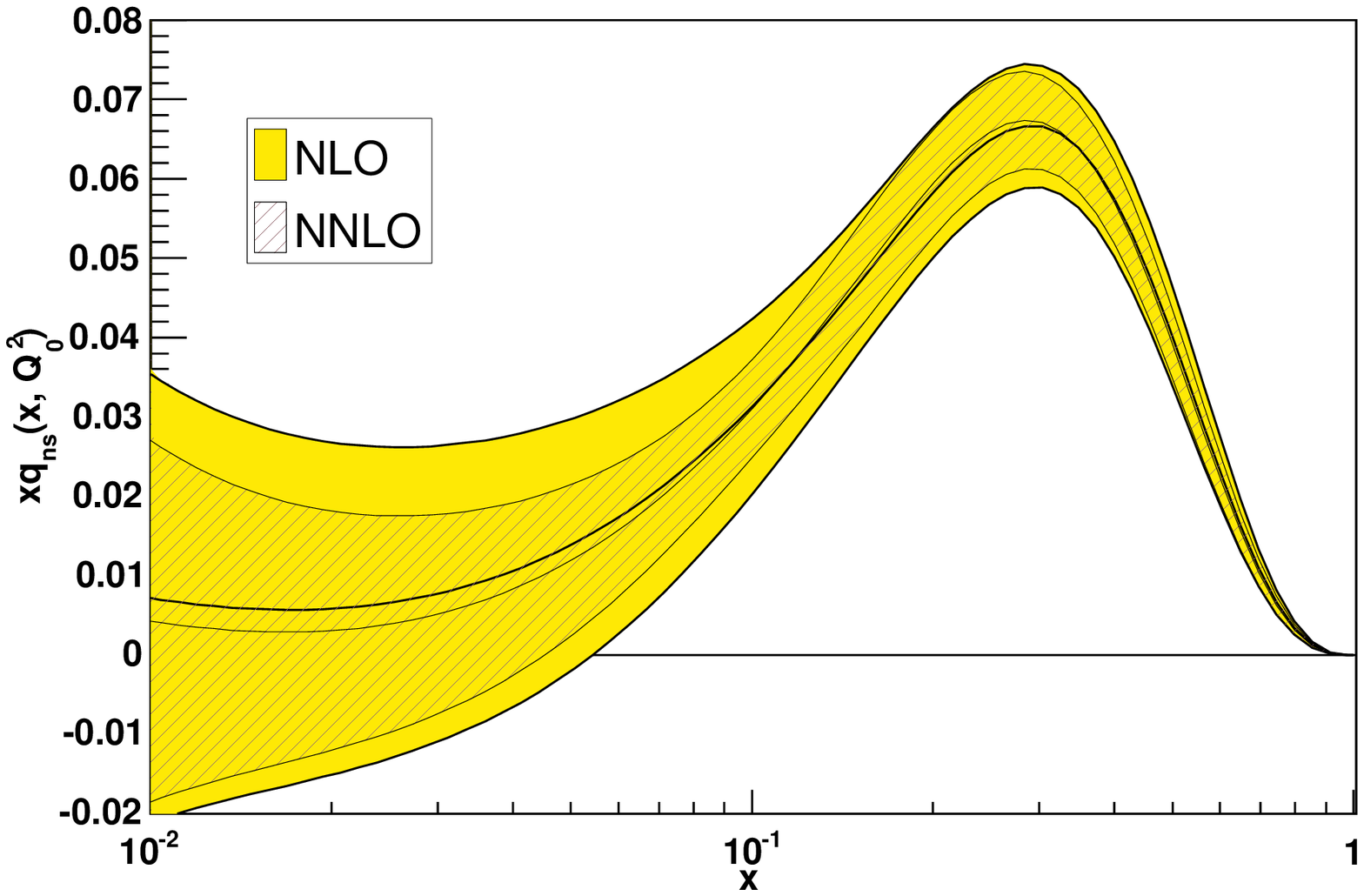}
\caption{Comparison of a LO fit (left) and a NNLO fit (right) to
the reference NLO fit.\label{order}}}


%
\FIGURE[ht]{
\epsfig{width=0.47\textwidth,figure=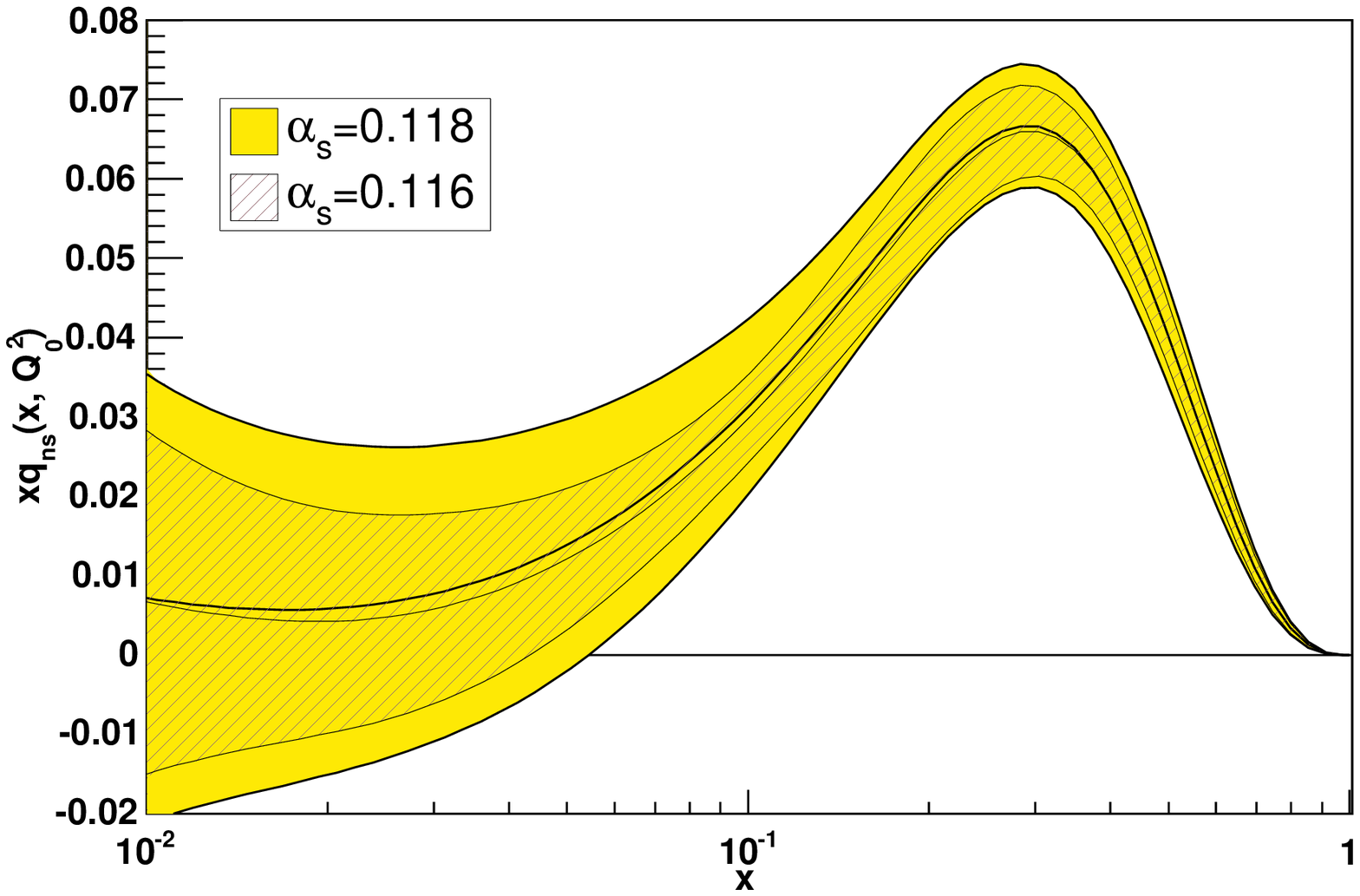}
\epsfig{width=0.47\textwidth,figure=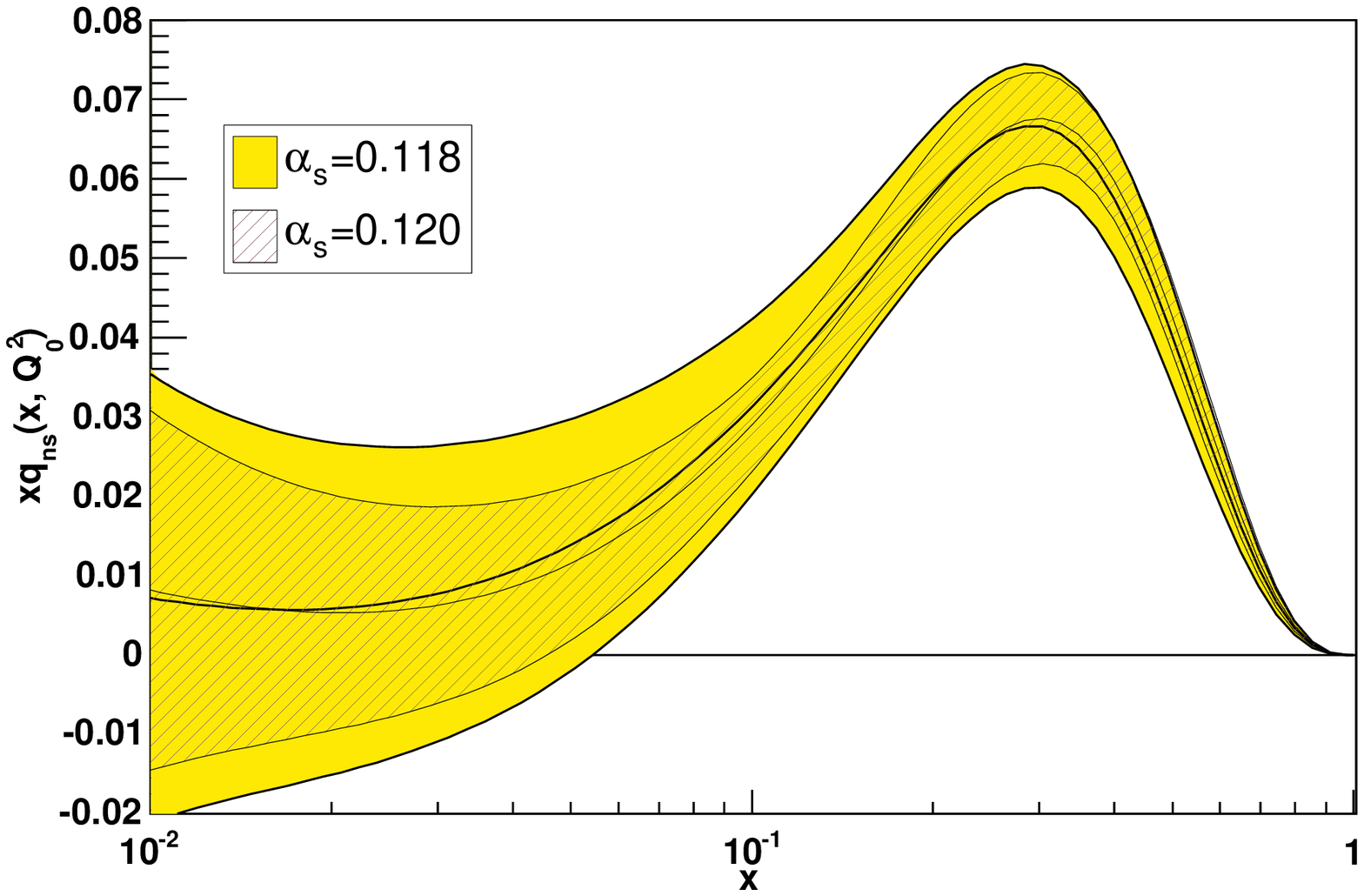}
\caption{Comparison of the reference fit to fits with a lower (left)
  or higher (right) value of $\amz$\label{alphafit}}
}
%

\TABLE{
\begin{tabular}{|c|c|c|c|}
\hline
$\alpha_s(M_Z^2)$  & $\qquad$ 0.116 $\qquad$  & $\qquad$  0.118 $\qquad$ 
 & $\qquad$  0.120 $\qquad$  \\
\hline
$\chi^2$ & 0.743 & 0.750 & 0.744   \\
\hline
$\la d\lc q \rc\ra_{\dat}$  & 3.8  & - &  4.1 \\
$\la d\lc q \rc\ra_{\extra}$  & 0.8 & - &  0.7\\
\hline
$\la d\lc \sigma_q \rc\ra_{\dat}$   & 1.6  & - & 2.4 \\
$\la d\lc \sigma_q \rc\ra_{\extra}$   & 1.4 & - & 1.5  \\
\hline
 $\la \sigma^{(\net)}
\ra_{\dat}$ &  $0.52~10^{-2}$   & $0.62~10^{-2}$&  $0.53~10^{-2}$  \\
\hline
\end{tabular}
\caption{\small Quantitative comparison of fits at NLO with different
values of $\alpha_s(M_Z^2)$. }
\label{alphastab}
}

The fact that NLO corrections have a negligible impact on the fit
suggests that the fit is only weakly dependent on  the value of
$\amz$. We have repeated our NLO fit while varying the value of
$\amz$ by one
sigma about its central value, i.e. with $\amz=0.116$ and  $\amz=0.120$. 
Comparison to the central fit (Fig.~\ref{alphafit} and
Tab.~\ref{alphastab})  shows that
the variation of the central fit 
due to the change in $\amz$ in this range has marginal statistical
significance: $\la d\lc q \rc\ra_{\dat}\sqrt{N_\dat}\approx 0.2$,
i.e., the central fit moves by about a fifth of a standard deviation
when $\amz$ is varied in this range. This variation, though small,
appears to be stable: if 50 replicas are used we get $\la d\lc q
\rc\ra_{\dat}=1.3$ for $\amz=0.116$ and $\la d\lc q
\rc\ra_{\dat}=1.1$ for $\amz=0.120$, corresponding to the same value 
of $\la d\lc q \rc\ra_{\dat}\sqrt{N_\dat}\approx 0.2$.
The variation of the $\chi^2$ in this range is also marginal and it
does not show a clear trend. We conclude that a NLO 
determination of $\alpha_s$ based on
nonsinglet data only is in principle possible, but it is necessarily 
affected by an uncertainty which
is  sizably larger than the error 
$0.002$ on the PDG average. A smaller value for this uncertainty 
would be  a likely indication of an underestimate of
the uncertainty related to the choice of parton parametrization. 
\section{The structure function}
\label{comp}

We finally turn to results for the physically measurable structure
function $F_2^{\rm \NS}(x, Q^2)$, which we compare to the data and to
results obtained in various other approaches.
\subsection{Comparison to data and other parton fits}
\label{compdata}

\FIGURE[ht]{\epsfig{width=0.75\textwidth,figure=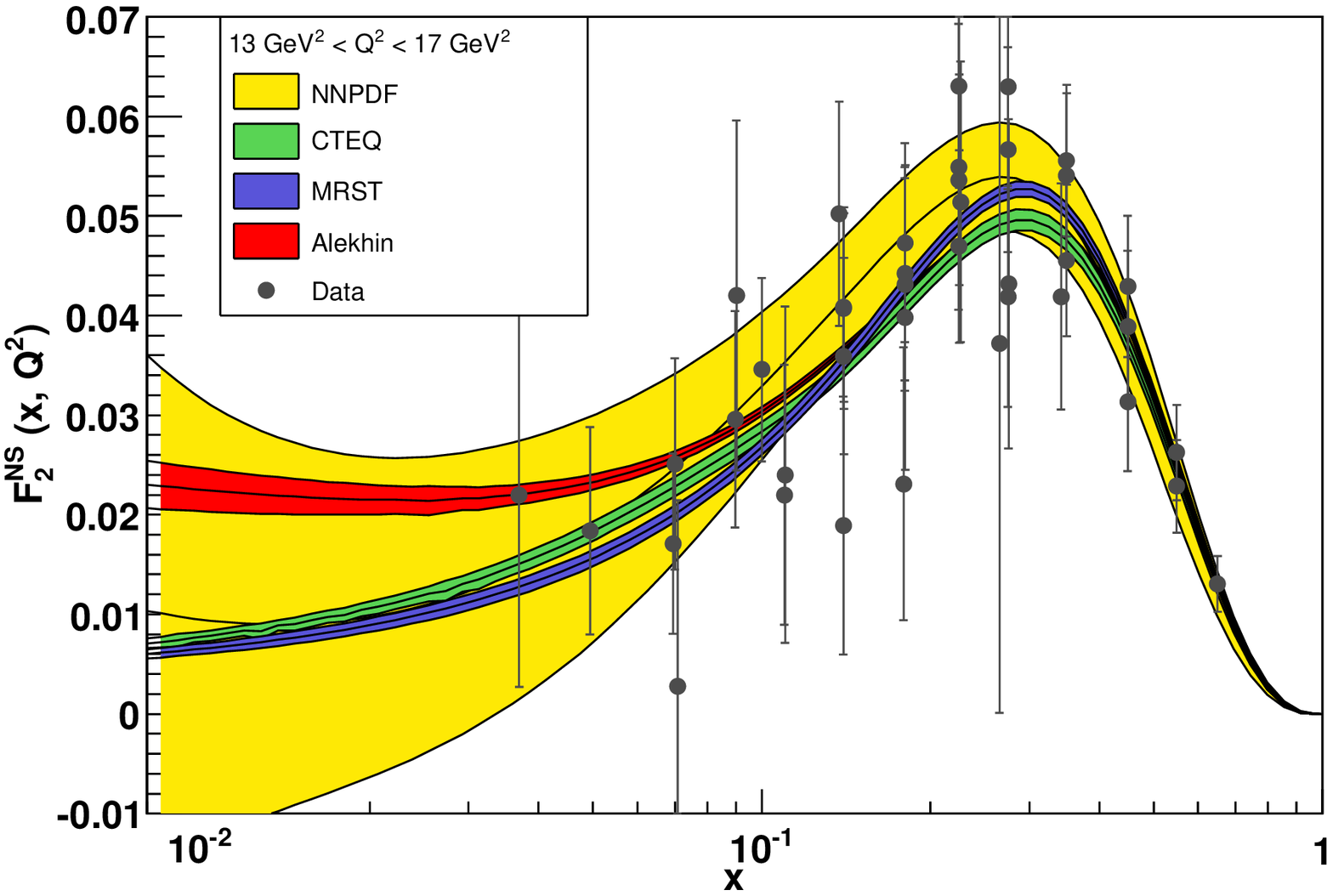}
\caption{Our NLO determination of 
the nonsinglet structure function $F_2^{\NS}(x,Q^2)$
at fixed $Q^2=15$~GeV$^2$ compared to all data with
$13 \le Q^2 \le 17$~GeV$^2$. 
Determinations by other groups are also shown.\label{f2data}}
}

\FIGURE[ht]{
\epsfig{width=0.75\textwidth,figure=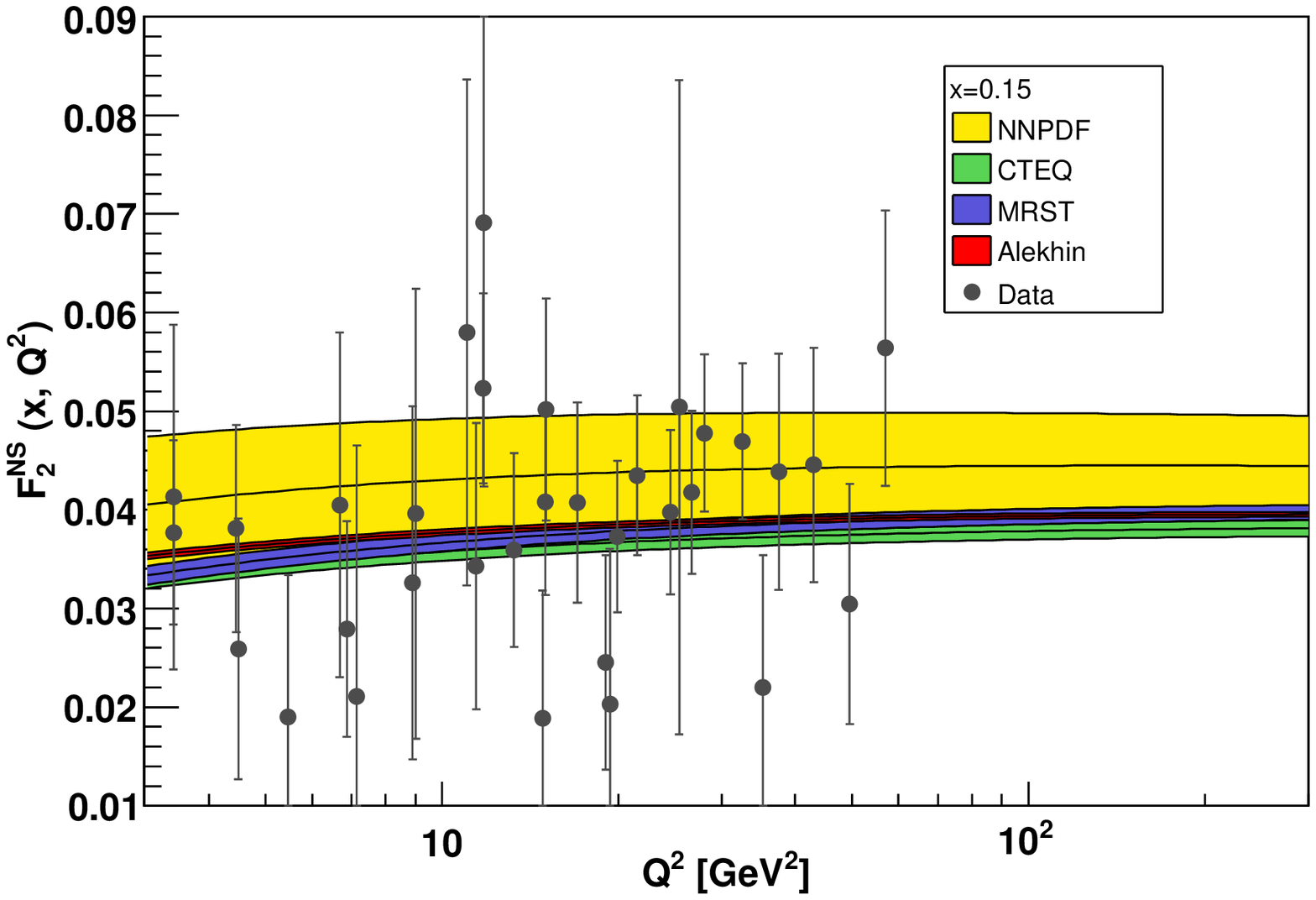}
\caption{Our NLO determination of 
the nonsinglet structure function $F_2^{\NS}(x,Q^2)$
as a function of
$Q^2$ for $x=0.15$, compared to all data 
in the range $0.13 \le x \le 0.17$. 
Determinations by other groups are also shown.\label{f2data2}}
}

In Figs.~\ref{f2data} and~\ref{f2data2} we show a comparison of our
results for the nonsinglet structure function $F_2^{\NS}(x,Q^2)$ as a
function of $x$ at fixed $Q^2=15$~GeV$^2$ compared to all data with
$13 \le Q^2\le 17$~GeV$^2$, and as a function of $Q^2$ at fixed
$x=0.15$, compared to all data in the range $0.13 \le x \le 0.17$. 
In the same figure, we
also show the structure function obtained from the results of the
global fits of the CTEQ~\cite{CTEQ} and MRST~\cite{MRST}
collaborations and the fit to deep-inelastic data by
Alekhin~\cite{Alekhin}.

The uncertainty in the final structure function is much smaller than
that on the data, thanks to the theoretical information from
perturbative evolution.  Nevertheless, the uncertainty in our
determination, especially at small $x$, is rather larger than that
found in previous analyses.  One may wonder whether this is due to the
fact that the global analyses are based on a wider set of
data. However, in the small $x$ region, where the difference between
the uncertainty on our result and that found by other groups is most
striking, none of the data contained in these global fits can
constrain $q_{\NS}(x,Q_0^2)$, on top of the data on $F_2^{\NS}(x,Q^2)$
which we also include. This suggests that the smaller error band
obtained by these groups may be due to parametrization bias or
uncertainty underestimation.

Be that as it may, wheras all available fits are within our error
band, our central result disagrees with these fits, which in turn
disagree with each other within respective errors, even in the valence
region $0.1\le x\le 0.3$.

\FIGURE[ht]{\epsfig{width=0.60\textwidth,figure=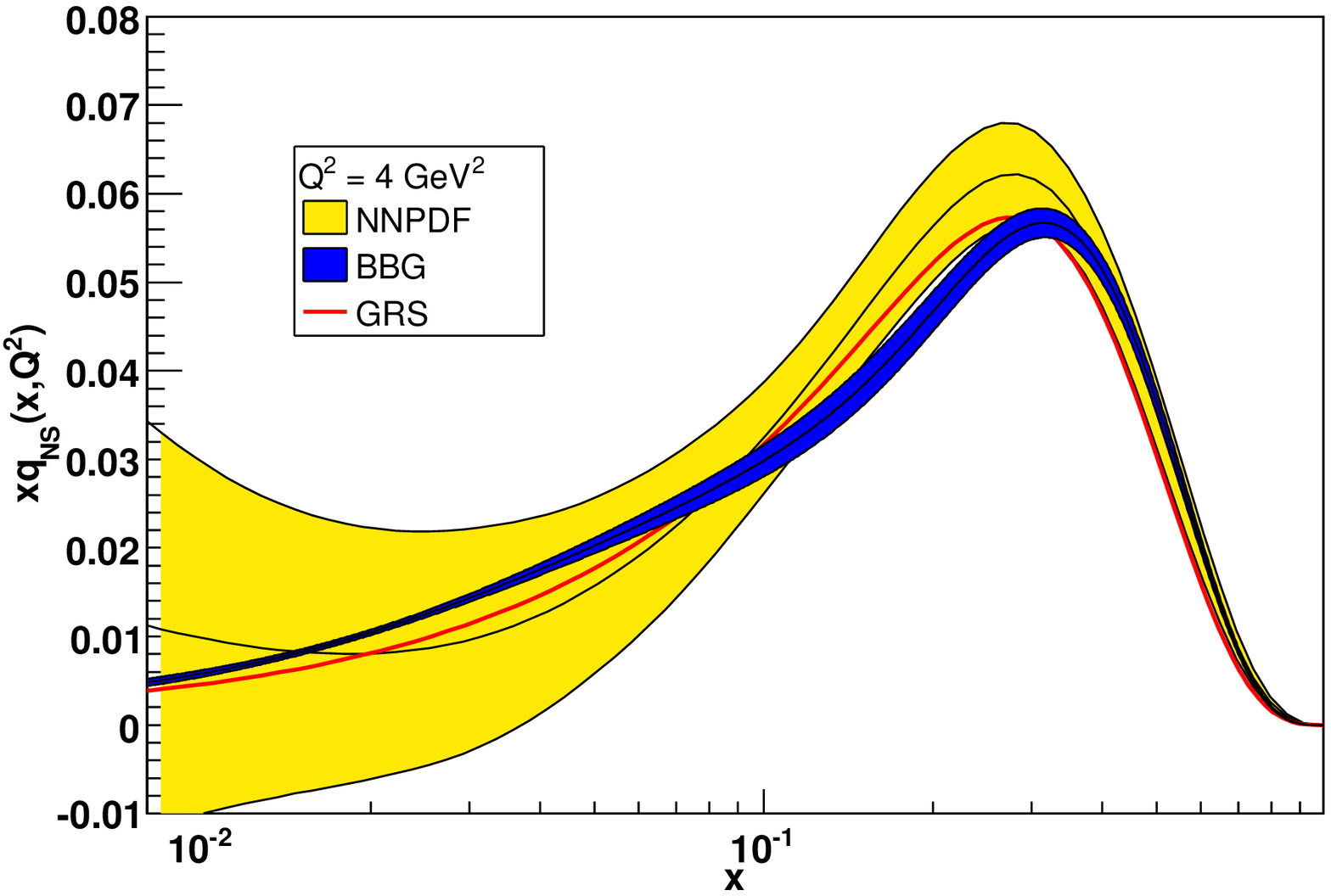}
\caption{Comparison of our NNLO fit with
the results of the nonsinglet fits of
Refs. \cite{bbg} and \cite{grs}. \label{nsfits}}}

Finally, it is interesting to compare our NNLO determination 
of the nonsinglet parton
distribution to the results of some 
recent NNLO fits to nonsinglet DIS data~\cite{grs,bbg} (see Fig.~\ref{nsfits}).
Results are analogous to those found when comparing to global fits.

\subsection{Comparison to neural $F_2^{\NS}(x,Q^2)$}

In Ref.~\cite{f2ns} the nonsinglet structure function
$F_2^{\NS}(x,Q^2)$ was directly parametrized with neural networks as a
function of its two arguments, instead of being obtained by evolving a
parton distribution. The results of that analysis are compared to the
present NLO determination in Fig.~\ref{compf2nn}. The comparison shows
that the uncertainty on the determination of Ref.~\cite{f2ns} is
already smaller than that of the data, because of the requirement of
smoothness of the function which is fitted to them, as discussed in
detail in Ref.~\cite{f2ns}. However, thanks to the extra information
from perturbative evolution, the uncertainty is further reduced by a
considerable amount when fitting the structure function.  Also, it is
apparent that thanks to this error reduction, a more detailed
determination of the shape of $F_2^{\NS}$ is possible.  We also
display the pull (see Appendix B) between these two determinations,
which shows that they are in perfect agreement within one sigma in the
region where there are data: whenever the pull increases in modulo
above one, no data are available.

\FIGURE[ht]{\epsfig{width=0.47\textwidth,figure=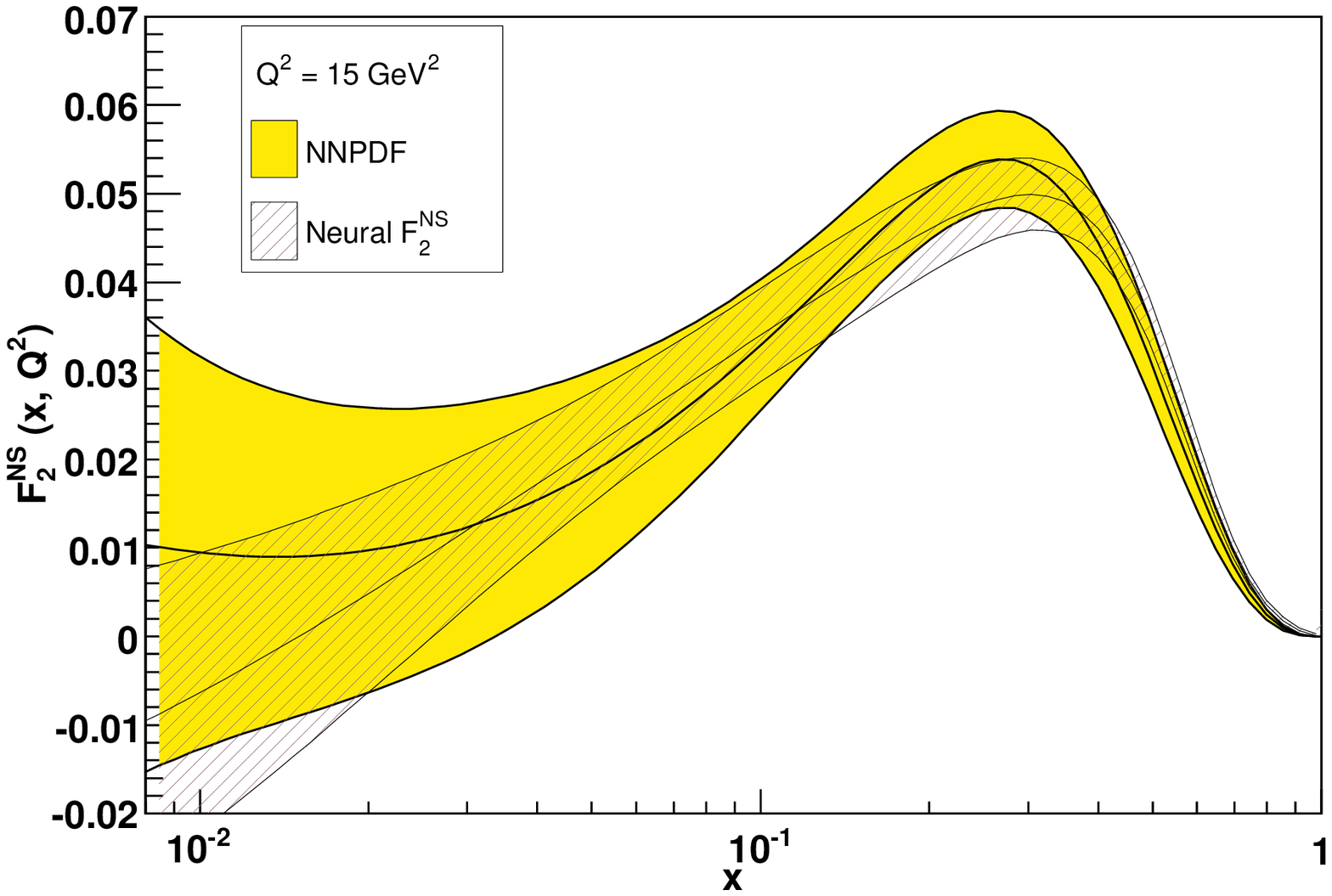}
\epsfig{width=0.47\textwidth,figure=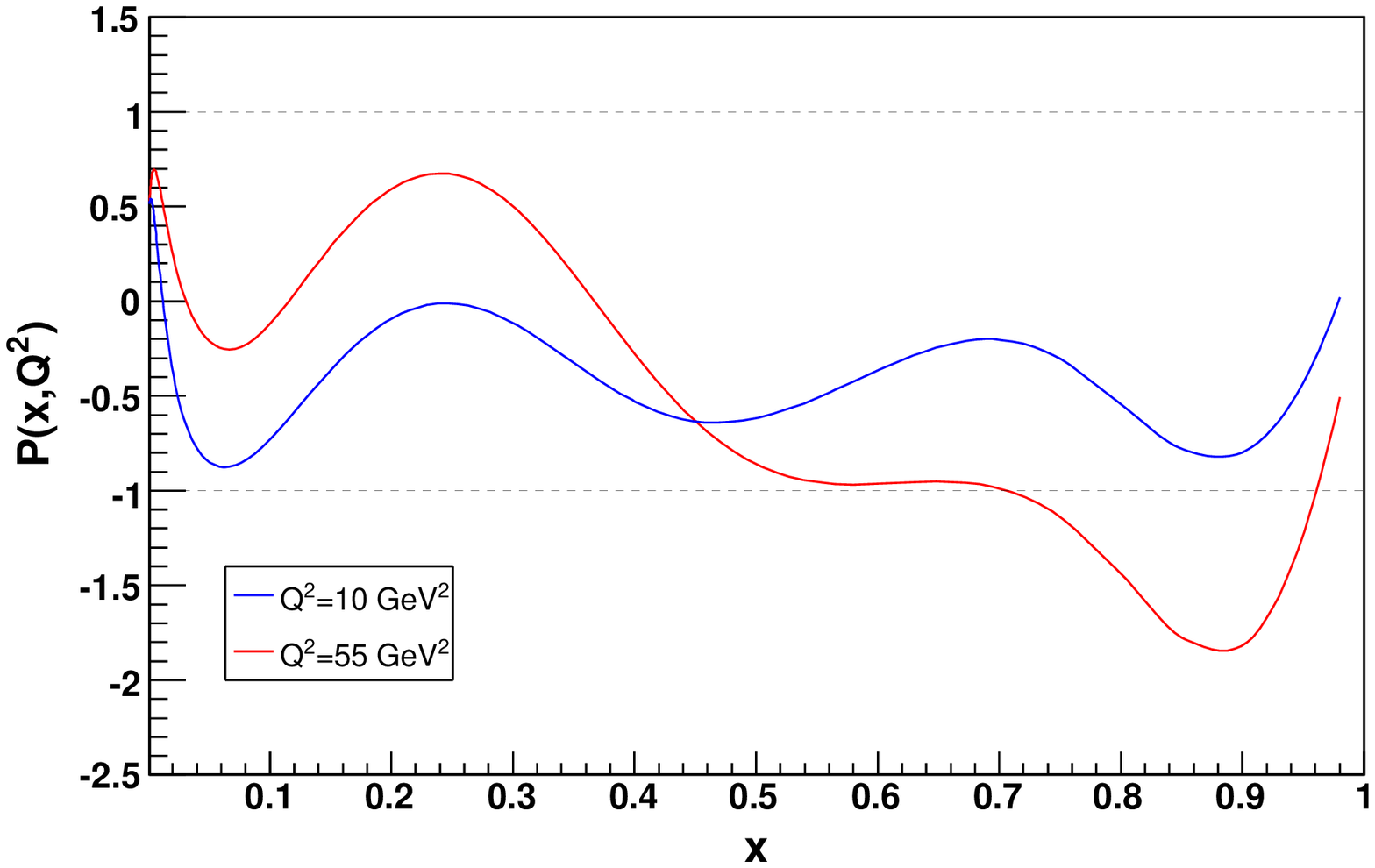}
 \caption{Comparison of the 
structure function $F_2^{\NS}(x,Q^2)$
          obtained from our NLO
$q_{\NS}(x,Q_0^2)$ with the NNPDF parametrization of
$F_2^{\NS}(x,Q^2)$ at $Q^2=15$ GeV$^2$: structure function (left) and
pull (right). The pull at $Q^2=55$ GeV$^2$ is also shown (lower curve
at large $x$).
}\label{compf2nn}}

\section{Conclusions and outlook}
In this paper we have provided a first determination of a parton
distribution within an approach based on the use of neural networks
coupled to a Monte Carlo representation of the probability density
in the space of structure functions.

Our final results take the form of a Monte Carlo sample of neural
networks.  The nonsinglet parton distribution $q_{\NS}(x,Q_0^2)$ and
its statistical moments (errors, correlations and so on) can be
determined by averaging over this sample. These results together with
driver FORTRAN code can be downloaded from the web site {\tt \bf
http://sophia.ecm.ub.es/nnpdf/}.

The main distinctive features of our determination of the isotriplet
quark distribution when compared to that obtained in other fits are
the following:
\begin{itemize}
\item Our fit is demonstrably independent of the
  parametrization, and the  uncertainty on it  has
  the correct statistical behavior which one expects of an unbiased
  statistical estimator.
\item The uncertainty is greatly reduced in comparison to that on the
  data, and also to that obtained in fits~\cite{f2ns} which do not
  exploit the constraint of perturbative parton evolution.
\item The error band on our result, especially at small $x$, is rather
  larger than that of other fits, and in fact it includes these other
  fits even when they disagree with each other. 
\item The quality of fits at LO, NLO and NNLO is the same: difference
  in perturbative order can be reabsorbed in the choice of boundary
  condition.
\end{itemize}

Our approach is designed to tackle some of the most serious problems
that plague current global parton fits, namely, the difficulty of
keeping under control the bias due to the choice of parametrization,
problems in combining different data sets, and the problem of
determining faithfully the uncertainty on the final result. This first
application to the determination of a single parton distribution from
all structure function data shows that the approach is viable and
successful. Its ultimate success or failure will be in its application
to a global parton fit from all available data.

\acknowledgments
This work was partly supported by grants ANR-05-JCJC-0046-01 (France),
PRIN-2004 (Italy), 
MEC FIS2004-05639-C02-01 and AP2002-2415
(Spain).
We thank R.~D.~Ball, M.~Cacciari, G.~d'Agostini, A.~Guffanti,
L.~Magnea, G.~Ridolfi 
and G.~Salam  for discussions during the course of this work.
\appendix

\section{Large $x$ behavior of the evolution factor}

At leading order in $\aq$, in the
large $x$ limit, the dominant contribution to the
evolution kernel comes from the large $N$ limit of the
LO anomalous dimension, given by  
\be
\label{largeN}
\Gamma(x)=\int_{-i\infty}^{+i\infty}\frac{dN}{2\pi i}x^{-N}
\lp \frac{\aq}{\aqq}\rp^{C_F\ln N /\pi\beta_0} .
\ee
The  inverse Mellin transformation can be performed to all logarithmic
orders 
using the formulas of Ref.~\cite{resum}, with the result
\bea
\label{largexgam}
\Gamma(x)&=&\sum_{n=1}^{\infty} \frac{\Delta^{(n-1)}(1)}{(n-1)!}
\lc \frac{1}{1-x} \frac{d^n}{d\ln^n(1-x)} \lp\frac{\aq}{\aqq}
\rp^{C_F\ln(1-x)/\pi\beta_0}\rc_++\mathcal{O}((1-x)^0) \nonumber \\
&=&\Delta\lp \frac{C_F}{\pi\beta_0}
\ln \lp\frac{\aqq}{\aq}
\rp \rp \lc (1-x)^{-1+
\frac{C_F}{\beta_0\pi}
\ln \lp\frac{\aqq}{\aq}
\rp }\rc_+ +\mathcal{O}((1-x)^0)\ .
\eea
where $\Delta(\eta)=1/\Gamma(\eta)$.
If $Q^2>Q_0^2$ the exponent of $1-x$ in
Eq.~(\ref{largexgam}) is larger than one, thereby ensuring that,
thanks to the
$+$ prescription, $\Gamma(x)$ is integrable. This in particular
implies that  all
integrals in Eq.~(\ref{evolfull}) exist.
Note  that in the
$Q^2\to\infty$ limit the $\Delta$ prefactor vanishes, consistent with
asymptotic freedom. 

\section{Statistical estimators}
\label{dataest}

We define  statistical estimators that are
used to asses the quality of both the Monte Carlo replica generation
(replica averages) and
the neural network training (neural network averages and statistical distance).
The superscripts $(\dat)$, $(\art)$ and $(\net)$ refer respectively to
the original data, to the $N_{\rep}$ Monte Carlo replicas of the data,
and to the  $N_{\rep}$  neural networks.
The subscripts $\rep$ and $\dat$ refer respectively to whether averages 
are taken  by summing over all replicas or over all data.

\begin{itemize}
\item{\bf Replica averages}
\begin{itemize}
\item Average over the number of replicas for each experimental point
  $i$
\be
\la
 F_i^{(\art)}\ra_{\rep}=\frac{1}{N_{\rep}}\sum_{k=1}^{N_{\rep}}
F_i^{(\art)(k)}\ .
\ee 
\item Associated variance
\be
\label{var}
\sigma_i^{(\art)}=\sqrt{\frac{N_\rep}{N_\rep-1}\left(\la\lp F_i^{(\art)}\rp^2\ra_{\rep}-
\la F_i^{(\art)}\ra^2_{\rep}\right)} \ .
\ee
\item Associated covariance
\be
\label{ro}
\rho_{ij}^{(\art)}=\frac{N_\rep}{N_\rep-1}\frac{\la F_i^{(\art)}F_j^{(\art)}\ra_{\rep}-
\la F_i^{(\art)}\ra_{\rep}\la F_j^{(\art)}\ra_{\rep}}{\sigma_i^{(\art)}
\sigma_j^{(\art)}} \ .
\ee
\be
\label{cov}
\mathrm{cov}_{ij}^{(\art)}=\rho_{ij}^{(\art)}\sigma_i^{(\art)}
\sigma_j^{(\art)} \ .
\ee
\item Percentage error on central values 
over the
  $N_{\dat}$ data points.
\be
\la PE\lc\la F^{(\art)}\ra_{\rep}\rc\ra_{\dat}=
\frac{1}{N_{\dat}}\sum_{i=1}^{N_{\dat}}\lc\frac{ \la F_i^{(\art)}\ra_{\rep}-
F_i^{(\mrexp)}}{F_i^{(\mrexp)}}\rc \ .
\ee
We define analogously $\la PE\lc\la
\sigma^{(\art)}\ra_{\rep}\rc\ra_{\dat}$. 
\item Scatter correlation:
\be
r\lc F^{(\art)}\rc=\frac{\la F^{(\mrexp)}\la F^{(\art)}
\ra_{\rep}\ra_{\dat}-\la F^{(\mrexp)}\ra_{\dat}\la\la F^{(\art)}
\ra_{\rep}\ra_{\dat}}{\sigma_s^{(\mrexp)}\sigma_s^{(\art)}} \,
\ee
where the scatter variances are defined as
\be
\sigma_s^{(\mrexp)}=\sqrt{\la \lp F^{(\exp)}\rp^2\ra_{\dat}-
\lp \la  F^{(\exp)}\ra_{\dat}\rp^2} \,
\ee
\be
\sigma_s^{(\art)}=\sqrt{\la \lp \la F^{(\art)}\ra_{\rep}\rp^2\ra_{\dat}-
\lp \la  \la F^{(\art)}\ra_{\rep} \ra_{\dat}\rp^2} \ .
\ee 
We define analogously $r\lc\sigma^{(\art)}\rc$, $r\lc\rho^{(\art)}\rc$
and $r\lc\mathrm{cov}^{(\art)}\rc$. Note that the scatter correlation
and scatter variance are not related to the variance and correlation
Eqs.~(\ref{var})-(\ref{cov}).
\item Average variance:
\be
\la \sigma^{(\art)}\ra_{\dat}=\frac{1}{N_{\dat}}
\sum_{i=1}^{N_{\dat}}\sigma_i^{(\art)} \ .
\label{avvar}
\ee
We  define analogously $\la\rho^{(\art)}\ra_{\dat}$ and
$\la\mathrm{cov}^{(\art)}\ra_{\dat}$,  as well as the
corresponding experimental quantities.
\end{itemize}

\item{\bf Neural network averages}
\begin{itemize}
\item Average error over networks
\be
\la E
\ra=\frac{1}{N_{\rep}}\sum_{k=1}^{N_{\rep}}{E      }^{(k)} \ ,
\label{aver}
\ee
where ${E}^{(k)}$ is given by Eq.~(\ref{er3}).

\item Average over neural networks for each experimental point
  $i$
\be
\la\label{netave}
 F_i^{(\net)}\ra_{\rep}=\frac{1}{N_{\rep}}\sum_{k=1}^{N_{\rep}}
F_i^{(\net)(k)} ,
\ee 
where $F_i^{(\net)}$ is the value of the nonsinglet structure function
  computed using Eq.~(\ref{f2conv}) from the neural network
  parametrization of the nonsinglet quark distribution at the values
  of $x$ and $Q^2$ which correspond to the $i$-th data point
\item Associated variance
\be
\label{netvar}
\sigma_i^{(\net)}=\sqrt{\frac{N_\rep}{N_\rep-1}\left(\la\lp F_i^{(\net)}\rp^2\ra_{\rep}-
\la F_i^{(\net)}\ra^2_{\rep}\right)} \ .
\ee
\item Associated covariance
\be
\label{netro}
\rho_{ij}^{(\net)}=\frac{N_\rep}{N_\rep-1}\frac{\la F_i^{(\net)}F_j^{(\net)}\ra_{\rep}-
\la F_i^{(\net)}\ra_{\rep}\la F_j^{(\net)}\ra_{\rep}}{\sigma_i^{(\net)}
\sigma_j^{(\net)}} \ .
\ee
\be
\label{netcov}
\mathrm{cov}_{ij}^{(\net)}=\rho_{ij}^{(\net)}\sigma_i^{(\net)}
\sigma_j^{(\net)} \ .
\ee
The corresponding means over
the  $N_{\dat}$ data points are computed using the analogue of
Eq.~(\ref{avvar}).
\item Scatter correlation
\be
r\lc F^{(\net)}\rc=\frac{\la F^{(\mrexp)}\la F^{(\net)}
\ra_{\rep}\ra_{\dat}-\la F^{(\mrexp)}\ra_{\dat}\la\la F^{(\net)}
\ra_{\rep}\ra_{\dat}}{\sigma_s^{(\mrexp)}\sigma_s^{(\net)}} \ .
\label{sccnets}
\ee 
We define analogously $\la\rho^{(\net)}\ra_{\dat}$ and
$\la\mathrm{cov}^{(\net)}\ra_{\dat}$. 
\end{itemize}

\item{\bf Distance between samples}
\begin{itemize}
\item Distance between different sets of neural nets
\be
\label{dist}
\la d[q]\ra=\sqrt{\la\frac{\left(\la q_i\ra_\one-\la
    q_i\ra_\two\right)^2} {\sigma^2[q_i^\one]+\sigma^2[q_i^\two]}\ra_\dat
},
\ee
with
\bea
\label{partnetave}\la q_i\ra_\one
&=&\frac{1}{N^\one_{\rep}}\sum_{k=1}^{N^\one_{\rep}}
q_{ik}^\one \\
\sigma^2[q_i^\one]
&=&\frac{1}{N^\one_{\rep} (N^\one_{\rep}-1)}\sum_{k=1}^{N^\one_{\rep}}
\left(q_{ik}^\one-\la q_i\ra_\one\right)^2,
\eea
where $q_{ik}^\one$ is the value of the nonsinglet quark distribution
for the $i$-th data point computed for the $k$-th replica from a set
of $N^\one_{\rep}$ neural networks, and likewise $ q_{ik}^\two$ is the
value computed for the $k$-th replica from a second set of
$N^\two_{\rep}$ neural networks.

\item The distance between errors and correlations is defined using
  Eq.~(\ref{dist}), but with the average value Eq.~(\ref{partnetave})
  replaced by the variance
\be
\label{partnetvar}
\sigma_i^{(\net)}=\sqrt{\frac{N_\one}{N_\one-1}\left(\la\lp q_i\rp^2\ra_{\one}-
\la q_i\ra^2_{\one}\right)} ,
\ee
and similarly for the covariance. Note that the distance is just the
square root of the 
sample variance normalized to the variance of which it is an unbiased
estimator (see Ref.~\cite{cowan}, in particular for a discussion of
the variance of $d$). 
\end{itemize}
\item{\bf Pull}

If  $F^{(a)}_2(x,Q^2)$ and $F^{(b)}_2(x,Q^2)$ are two different
determinations of the structure function, each affected respectively
by uncertainty $\sigma_{(a)}(x,Q^2)$ and $\sigma_{(b)}(x,Q^2)$, the
pull is defined as  
\be
\label{pull}
\la P(x,Q^2)\ra=\frac{\left( F^{(a)}_2(x,Q^2)-
    F^{(b)}_2(x,Q^2)\right)}
{\sqrt{\sigma^2_{(a)}(x,Q^2)+\sigma_{(b)}^2(x,Q^2)}}.
\ee

\end{itemize}

\bibliography{nnqns-1.18}

\end{document}